\documentclass{jpp}
\usepackage{graphicx}

\usepackage[utf8]{inputenc}
\usepackage[T1]{fontenc}
\usepackage{amsmath}
\usepackage{url}
\usepackage{color}
\usepackage{lineno} 

\shorttitle{Confinement performance predictions for a high field tandem mirror}
\shortauthor{S. J. Frank}

\title{Confinement performance predictions for a high field axisymmetric tandem mirror}

\author{S. J. Frank\aff{1}
 J. Viola\aff{1},
 Yu. V. Petrov\aff{2},
 J. K. Anderson\aff{1},
 D. Bindl\aff{1},
 B. Biswas\aff{1},
 J. Caneses\aff{2},
 D. Endrizzi\aff{1},
 K. Furlong\aff{1},
 R. W. Harvey\aff{2},
 C. M. Jacobson\aff{1},
 B. Lindley\aff{1},
 E. Marriott\aff{1},
 O. Schmitz\aff{1},
 K. Shih\aff{1},
 D.A. Sutherland\aff{1},
 \and C. B. Forest\aff{1},
 }

\affiliation{\aff{1}Realta Fusion Inc., Madison, WI 53717, USA
\aff{2}CompX, Del Mar, CA 92014, USA}

\begin{document}
\maketitle

\begin{abstract}
This paper presents Hammir tandem mirror confinement performance analysis based on Realta Fusion's first-of-a-kind model for axisymmetric magnetic mirror fusion performance. This model uses an integrated end plug simulation model including, heating, equilibrium, and transport combined with a new formulation of the plasma operation contours (POPCONs) technique for the tandem mirror central cell. Using this model in concert with machine learning optimization techniques, it is shown that an end plug utilizing high temperature superconducting magnets and modern neutral beams enables a classical tandem mirror pilot plant producing a fusion gain Q > 5. The approach here represents an important advance in tandem mirror design. The high fidelity end plug model enables calculations of heating and transport in the highly non-Maxwellian end plug to be made more accurately. The detailed end plug modelling performed in this work has highlighted the importance of classical radial transport and neutral beam absorption efficiency on end plug viability. The central cell POPCON technique allows consideration of a wide range of parameters in the relatively simple near-Maxwellian central cell, facilitating the selection of more optimal central cell plasmas. These advances make it possible to find more conservative classical tandem mirror fusion pilot plant operating points with lower temperatures, neutral beam energies, and end plug performance requirements than designs in the literature. Despite being more conservative, it is shown that these operating points have sufficient confinement performance to serve as the basis of a viable fusion pilot plant provided that they can be stabilized against MHD and trapped particle modes.
\end{abstract}

\section{The compact mirror approach to fusion}\label{sec:intro}

Magnetic mirrors have several variants. This work will focus on the simple mirror and the "classical" tandem mirror. The simple mirror is a fusion configuration which confines plasma in a magnetic field well created by two large electromagnets using the magnetic mirror force, $\boldsymbol{F} =- \mu_m \bnabla B$ arising from the conservation of the magnetic moment $\mu_m = m v_\perp^2 / 2B$, where $v_\perp$ is the perpendicular component of a particle's velocity, $m$ is a particle's mass, and $B$ is the background magnetic field. The ratio of the magnetic field at the centre of the well $B_0$ and at the edge of the well in the centre of the electromagnet, sometimes referred to as the mirror throat $B_m$, is known as the mirror ratio $R_m = B_m / (B_0 \sqrt{1-\beta})$. The factor of $\sqrt{1-\beta}$, where $\beta = 2\mu_0 p / B_0^2$, in the mirror ratio equation comes from plasma diamagnetism. Particles with low enough parallel velocity, $v_\parallel \leq v_\perp \sqrt{R_m-1} $, remain trapped in the magnetic mirror and leak out through collisions (and in some cases kinetic instabilities generated by the plasma's distribution function). It is known that if high energy neutral beams are injected into a simple mirror particle confinement times $\tau_p \sim 1~\mathrm{s}$ are achievable in theory \citep{Post1987}, but in practice under realistic engineering and physics constraints it is unlikely that a simple mirror can generate a fusion gain $Q = P_{\mathrm{fus}}/P_{\mathrm{heat}} \gg 1$, where $P_{\mathrm{fus}}$ is the fusion power and $P_{\mathrm{heat}}$ is the external heating power \citep{Freidberg2007,Forest2024}. The classical tandem magnetic mirror was designed to overcome this limitation \citep{Dimov1976,Fowler1977,Dimov2005}. In a tandem mirror, a central cell consisting of a long solenoid is flanked by two simple mirror end plugs. Intense heating power is directed into the end plugs to create hot, dense, plasmas. This causes a large electrostatic potential drop between the end plugs and the lower density central cell. The potential difference confines an effectively Maxwellian ignited plasma in the central cell. Tandem mirrors can achieve very large confinement times in the central cell $\tau_p \sim 10~\mathrm{s}$, overall system gains much greater than unity, and operate in steady-state without the risk of destructive plasma disruptions. However, early analysis of classical tandem mirrors suggested that they would require very large ratios of end plug to central cell density, mirror ratios, and plasma temperatures that were outside of the capabilities of the technology available at the time \citep{Dimov1976,Fowler1977,Dimov2005}. Because of this, other options for enhancing tandem performance such as "thermal barriers" were pursued in historical tandem mirror designs until the eventual discontinuation of the US mirror development program \citep{Baldwin1979,Logan1983,Post1987}. A number of tandem mirror experiments, both with and without thermal barriers, have been constructed since their proposal, but none have yet reached fusion relevant parameter spaces \citep{TMXGroup1981,Porter1988,Brau1988,Dimov1993,Tamano1995,Akhmetov1999}

The new magnetic mirror development program being undertaken by Realta Fusion takes advantage of a number of key enabling technology and physics breakthroughs that can greatly enhance the performance of the magnetic mirror to revive the tandem mirror fusion plant concept: high-energy negative-ion neutral beams \citep{Kashiwagi2021,Serianni2023}, modern high-frequency gyrotrons \citep{Rzesnicki2022,Bridge12Gyro}, HTS magnets \citep{Michael2017,Whyte2019}, and  magnetohydrodynamic (MHD) stabilization of axisymmetric mirrors \citep{Beklemishev2010, Ryutov2011, Endrizzi2023}. These advancements allow less complex magnetic mirrors to be constructed with planar HTS coils that reach substantially higher fields and mirror ratios than previous magnetic mirror experiments and can dramatically improve tandem mirror performance \citep{Fowler2017}. Presently, Realta is sponsoring research at the University of Wisconsin-Madison on the Wisconsin High-Temperature Superconducting Axisymmetric Magnetic Mirror (WHAM) experiment, an axisymmetric mirror using two custom $17~\mathrm{T}$ HTS magnetic field coils built by Commonwealth Fusion Systems. A comprehensive description of the WHAM experiment is available in \citet{Endrizzi2023}. The next step device, Anvil, will be a large simple mirror constructed by Realta Fusion to serve as an end plug physics demonstrator and fusion technology test platform. Anvil will be a similar class of device to the break-even axisymmetric mirror (BEAM) described in \citet{Forest2024}. The Realta Fusion tandem mirror pilot plant Hammir is the final step on Realta Fusion's technology development roadmap and will be the focus of this paper. Hammir will be a smaller, more conservative, axisymmetric tandem mirror than those identified in the \citet{Fowler2017} analysis, which considered only axisymmetric tandem mirrors with $\beta \sim 1$, very high temperatures $T_i = T_e = 150~\mathrm{keV}$, high energy $1~\mathrm{MeV}$ neutral beams, and large fusion power outputs $P_{fus}>1~\mathrm{GW}$. It will be shown in this work that smaller classical tandem mirrors with significantly lower $\beta$, temperature, beam energies, and fusion power are still competitive with other fusion concepts and compatible with the requirements for a commercial fusion plant identified by the National Academies of Sciences, Engineering, and Medicine (NASEM) “Bringing Fusion to the U.S. Grid” report \citep{Hawryluk2021}, specifically:
\begin{itemize}
    \item Electrical gain: $Q_e = P_{\mathrm{ele,out}}/P_{\mathrm{ele,in}} > 1$
    \item Continuous net electricity $P_{\mathrm{ele,out}} > 50~\mathrm{MWe}$ for at least 3 hours (or the process heat equivalent in the case of a industrial process heat use case)
\end{itemize} 

To simplify tandem mirror modelling, this work takes advantage of the properties of the classical tandem mirror to split up the modelling problem. The central cell and the end plug are handled using separate models that are then coupled together. The complicated non-Maxwellian end plug which undergoes intense heating is analysed using a high fidelity integrated simulation model, whereas, the simpler near-Maxwellian central cell is extrapolated based on the end plug parameters using a novel version of the Plasma OPeration CONtour (POPCON) technique \citep{Cordey1981,Houlberg1982}. This approach allows larger parameter spaces to be considered and breaks the tandem mirror design problem down to a few key engineering and physics parameters which are important to design such as the plasma radius at the high field mirror coil throat $a_m$, the field at the mirror throat $B_m$, and the end plug density $n_p$. These are key steps to developing effective pilot plant designs, and successful tokamak designs also have used multifidelity models and identified analogous design parameters such as the minor radius $a$, major radius $R$, on axis field $B_0$, and plasma current $I_p$ \citep{Sorbom2015,Creely2020,Buttery2021,Frank2022,Rutherford2024}. While this work considers stability against some kinetic and MHD instabilities, it does not include a comprehensive evaluation of stability. The RealTwin integrated simulation framework developed here for equilibrium, transport, and heating prediction is intended to serve as a platform for future simulations of stability based on realistic equilibria and plasma distribution functions. A demonstration of WHAM simulations of transport and equilibrium coupled to stability simulations using this framework can be found in the companion article \citet{Tran2024}. Such simulations will be extended to the tandem mirror analyses here in future work. 

The rest of this paper is structured as follows: Section~\ref{sec:computational_model} describes the new RealTwin$^{\mathrm{TM}}$ integrated model for a Hammir end plug. Section~\ref{sec:POPCONs} determines what end plug parameters are required for a viable Hammir fusion power plant using a central cell POPCON. Section~\ref{sec:optimization} identifies an optimal operating point for the Hammir end plug which satisfies the end plug requirements identified by the POPCONs in Section~\ref{sec:POPCONs} under realistic physics and engineering constraints using the RealTwin model. Finally, Section~\ref{sec:conclusion} summarizes the work, identifying key capability gaps discovered during the Hammir design process and plans to address them in the future. 

\section{The RealTwin simulation model}\label{sec:computational_model}
\begin{figure*}
  \centering
  \includegraphics[width=1\linewidth]{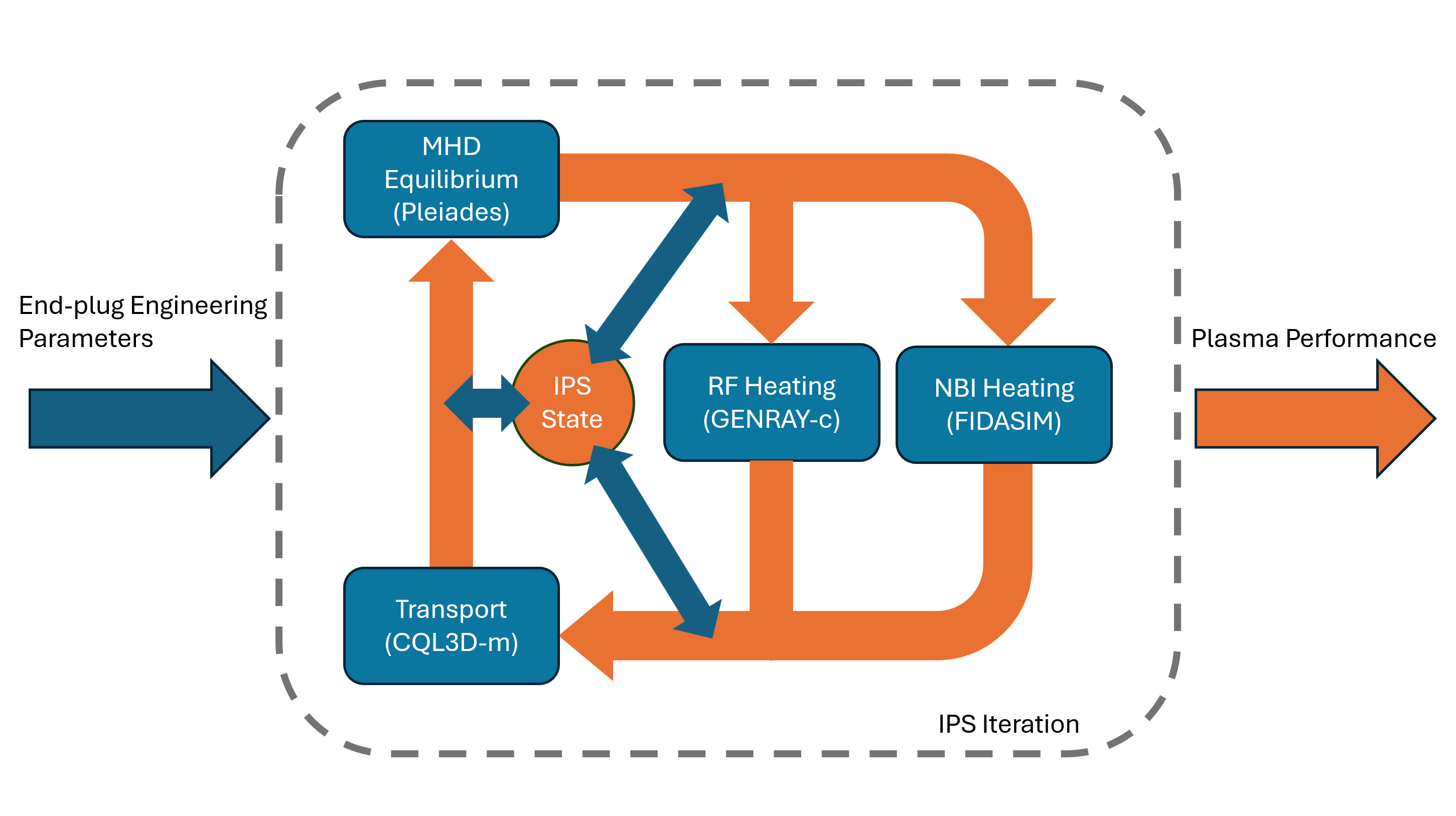}
  \caption{An example of an IPS driven workflow for calculating simple mirror plasma performance. After each code in the iteration is run the plasma state is updated and used to setup the next iteration.}
\label{fig:ips_workflow}
\end{figure*}

The RealTwin uses an integrated simulation model including Fokker-Planck transport, anisotropic magnetic equilibria as well as heating from neutral beams and radiofrequency (RF) waves. Detailed descriptions of the Fokker-Planck model, CQL3D-m and the equilibrium model, Pleiades will be given later in this section. In order to couple simulations together in a self-consistent fashion, a modified version of the Integrated Plasma Simulator (IPS) \citep{Elwasif2010} was created with wrappers specifically tailored to mirror relevant codes and an HDF5 based mirror plasma state file. The IPS is designed to enable many codes to be run in a single workflow using a driver program that coordinates the execution of different codes, manages their resource usage, and supports advanced features such as concurrent and nested execution of simulations. When each code runs, it interfaces with the IPS through the centralized plasma state file allowing different codes and models to be swapped in and out of the workflow in a modular fashion. This has made the IPS the tool of choice for many public fusion research initiatives including ARPA-E projects and SciDACs \citep{Bonoli2018, Badalassi2023, Collins2023} and has driven large scale coupled simulations using CQL3D previously \citep{Frank2022_JPP,Frank2024}. Usage of the IPS enables many different physics codes to be coupled together in an efficient manner enabling \textit{integrated} simulation of mirror plasmas in which the right code can be used for the right job. It also facilitates the level of repeatability and automation needed to do optimizations like the one presented here.

An IPS simulation is initialized with defeatured engineering information about the mirror system under consideration. From that information, the IPS generates a plasma state file which contains the heating system parameters, the plasma parameters, and the magnetic coil parameters as well as the magnetic equilibrium. Transport simulations, like the ones reported on here, are initialized assuming a uniform breakdown plasma (the parameters used for the initial plasma have not been found to notably affect the simulations' results at steady-state with the exception of the situation in which the initial density is too low to provide significant NBI absorption). The first simulation executed is generally Pleiades to provide a calculation of the vacuum magnetic fields for a given set of coil parameters. Heating sources from RF and/or neutral beams (if an external neutral beam code like FIDASIM is used rather than the Freya solver in CQL3D-m) are then computed, and a transport simulation with CQL3D-m is run for $\tau \lesssim \tau_p \sim 1~\mathrm{s}$. The updated distribution functions computed by CQL3D-m are integrated to provide anisotropic pressure profiles which are used to rerun Pleiades to provide an updated magnetic equilibrium and the simulation loop is restarted. This loop continues, typically for ten or more iterations until a converged equilibrium solution is obtained. An entire simulation of this type takes between 12 to 24 hours of wall-clock time running on two nodes on the NERSC Perlmutter supercomputer. An example of an IPS workflow similar to the one described here is shown in Figure~\ref{fig:ips_workflow}.

To optimize designs over large multidimensional engineering parameter spaces, Bayesian optimization (BO) using a Gaussian process (GP) has been applied to IPS simulations. In this technique, a GP is used to construct a surrogate model which replicates the behaviour objective function which optimizes some quantity based on the RealTwin integrated model using data points obtained by running the model. This objective function surrogate model serves as the BO's prior. An acquisition function is then used to determine the optimal points to sample to reduce uncertainty in the objective function and optimize it. Similar optimization procedures have been applied tokamaks for both single objective and multi-objective problems \citep{Nunn2023,Pavone2023,Brown2024,Rodriguez2024}. The SciKit Bayesian optimization library \citep{scikit-learn} was used here to scan over the engineering parameters while optimizing for end-plug performance. This work used a single objective function modified by cost functions to avoid parameter regimes approaching instabilities (further details on these cost functions will be given in Section~\ref{sec:optimization}). Future work plans to utilize a more sophisticated multi-objective optimization.

Given the wall-clock expense for each RealTwin simulation, it makes the most sense to optimize in batches. At the start of each batch, an acquisition function is used to generate multiple points to sample using RealTwin simulations. These points are input into simulations using an automated workflow and submitted to a supercomputer to run. Once a batch is completed the results of that batch are added to the existing results to reduce the error in the BO's GP prior and the acquisition function is recalculated. Sampling is performed using a Constant Liar "CL-min" \citep{chevalier2013fast} approach with a value of $\kappa$ between 5 and 10 (lower values of $\kappa$ indicate less exploration and more optimization). A global optimum is typically found within 40-60 RealTwin runs (4-6 batches with 10 samples per batch). 

While not included in the simulations presented here, a number of additional codes are under development by Realta Fusion and collaborators which are in varying stages of IPS integration. In particular, significant progress has been made with integration of Hybrid-VPIC \citep{Le2023} simulations into the IPS workflow such that ion-scale kinetic stability and MHD stability of the plasma equilibria computed by the IPS workflow can be evaluated. The CQL3D-m distribution functions and the equilibrium magnetic fields computed by Pleiades are used to initialize a Hybrid-VPIC simulation in these cases, and such simulations are presented in this publication's companion article \citet{Tran2024}. In principle, if the spatial and velocity diffusion computed by Hybrid-VPIC can be mapped back into CQL3D-m using the IPS, it could provide a full classical \textit{and} kinetic transport model for a simple mirror (as well as an assessment of MHD stability). This method of integrated transport modelling is similar to the integrated transport models used in tokamaks such as TGYRO/NEO/CGYRO \citep{Belli2008,Belli2011,Candy2009,Candy2016}. Edge neutrals modelling is also not included in the present model. Ingress of neutrals into the mirror plasma and their charge exchange with mirror trapped fast ions can be an important loss mechanism in small magnetic mirror experiments, and it will be necessary to include this effect to achieve quantitative simulation agreement in WHAM. Also underway is tighter integration of the IPS driven RealTwin simulations with Realta Fusion's engineering design models. This integration will include features such as checks on component interference during IPS simulation setup and the ability to import IPS driven simulation results to form fusion neutron sources in neutronics simulations.

\subsection{Fokker-Planck transport and plasma heating using CQL3D-m and Freya}

Mirror transport during high-performance operation where MHD and trapped particle flute-like modes as well as kinetic modes such as the drift-cyclotron loss-cone (DCLC) instability are not present is dominated by axial transport from collisions and neutral dynamics. To simulate transport under these conditions, a Fokker-Planck equation solver including plasma heating and fuelling models is employed. This work used the CQL3D-m Fokker-Planck code and its internal neutral beam deposition Monte-Carlo solver Freya. 

The CQL3D-m code \citep{Harvey2016} used here is a modified version of the well known CQL3D Fokker-Planck code \citep{Harvey1992} and older versions of CQL3D-m were used for the design of WHAM and BEAM devices \citep{Endrizzi2023,Forest2024}. Much like CQL3D, CQL3D-m solves the bounce-averaged Fokker-Planck equations for a coupled set of plasma species $\alpha, \beta, ...$. The Fokker-Planck equation for plasma species $\alpha$ is:
\begin{equation} \label{eq:fokker-planck}
    \frac{df_{\alpha 0}}{dt} = C(f_{\alpha 0},f_{\beta 0},...) + Q(f_{\alpha 0},\boldsymbol{E}) + R(f_{\alpha 0}) + S_{+}(f_{\alpha 0}) - S_{-}(f_{\alpha 0}),
\end{equation}
where $f_{\alpha 0}$ is the bounce-averaged distribution function of species $\alpha$ at the midplane, where $z=0$ (values taken at the midplane are denoted with subscript $0$), $f_{\beta 0}$ is the distribution function of species $\beta$ etc., $C(f_{\alpha 0},f_{\beta 0},...)$ is the bounce-averaged nonlinear form of the Rosenbluth collision operator \citep{Rosenbluth1957} considering collisions between all species in the simulation using their fully numerical distribution functions, $Q(f_{\alpha 0}, \boldsymbol{E}_{RF})$ is the bounce-averaged quasilinear diffusion coefficient associated with radiofrequency (RF) electric field $\boldsymbol{E}$ arising from either RF heating or an instability such as the DCLC, $R(f_{\alpha 0})$ is the radial diffusion (not considered in this work as losses from radial diffusion are ideally much smaller than axial losses in the end plug), and $S_+$ and $S_-$ are particle source and sink terms respectively arising from mirror losses, neutral fuelling, and charge-exchange. 

CQL3D-m discretizes \eqref{eq:fokker-planck} in two velocity-space dimensions, $u = \gamma v$ the momentum-per-rest-mass (where $v$ is particle velocity and $\gamma$ is the relativistic factor) and $\vartheta$ the velocity-space pitch angle, as well as one spatial dimension $\psi$ the square-root poloidal flux (a radial-like coordinate in the magnetic mirror), and uses the implicit time advance solver described in \citet{Kerbel1985}. The coefficients $Q$, $C$, $R$, $S_\pm$ in \eqref{eq:fokker-planck} have been bounce-averaged in magnetic mirror geometry assuming zero-orbit-width. The bounce-average of some coefficient $A$ is defined:
\begin{equation}
    \langle \langle A \rangle \rangle = \frac{1}{\tau_b}\oint \frac{d\ell_b}{\left|v_\parallel\right|} A \Bigg|_{\mu_m, \vartheta_0, \psi} .
\end{equation}
and is performed over the equilibrium fields (vacuum fields are only used in the first iteration between CQL3D-m and Pleiades).

The mirror loss boundaries in CQL3D-m are calculated by the following method. Particles starting with a certain $\vartheta_0$ and $u_0$ at the midplane can be trapped by either the magnetic mirror force or the mirror's ambipolar potential. These two trapping mechanisms have distinct loss boundaries that can be computed by the following methods. Potential trapping occurs when for species $\alpha$, the quantity $v_p^2(z)$ becomes less than zero at some point in $z$ along the flux surface. The definition of this quantity is:
\begin{equation}
    v_p^2(z) = 2\frac{q_\alpha}{m_\alpha} (\phi_{p0} - \phi_p(z)), 
\end{equation}
where $\phi_p(z)$ is the ambipolar potential \citep{Pastukhov1974,Cohen1978,Khudik1997}. This can be used to define a potential trapping boundary where:
\begin{equation} \label{eq:pot_bound}
    v_{\phi_p}(z) = \sqrt{-v_p(z)^2}.     
\end{equation}
For a species where $v_p^2 > 0$ along the entire flux surface in $z$, no potential trapping takes place. Similarly, the mirror trapping boundary for species $\alpha$ with $v_0$ can be computed with equations:  
\begin{subequations} \label{eq:mirror_bound}
\begin{align}
    v_{\parallel , \mathrm{mirr}} &= v_{0} \sin{\vartheta_{\mathrm{mirr}}} \\
    \vartheta_{\mathrm{mirr}} &= \arcsin{\left(\sqrt{\frac{B(z)}{B_0} \frac{v_0^2}{v_0^2 + v_p(z)^2}}\right)_{\mathrm{max},z}},
\end{align}
\end{subequations}
where the quantity in parentheses is evaluated at its maximum value in $z$ along a poloidal flux surface. If $\phi_p(z)$ is taken to be zero everywhere, this becomes the standard expression for the mirror loss cone. Components of the distribution function $f$ which have velocities less than the outermost loss boundary are removed by a sink term:
\begin{equation}
    S_-^{loss}(f) = \frac{2v_\parallel}{L}f  
\end{equation}
where $L$ is the length of the mirror from mirror throat to mirror throat. The loss boundaries are then recomputed for the next timestep based on an updated $\phi_p(z)$ profile calculated with the method below

CQL3D-m accounts for ambipolarity preservation by directly calculating the ambipolar potential response \citep{Pastukhov1974,Cohen1978,Khudik1997}. This is done through an iterative relaxation solver which determines the potential required to meet the quasineutrality condition $n_e = \sum_{si} Z_{si} n_{si}$. The ambipolar potential at a given $z$ location along a flux surface is computed by iterating two different calculation methods. The first method is an iterative solver which takes:
\begin{equation} \label{eq:method1}
    \phi_p^{(n)}(z) = \phi_p^{(n-1)}(z) + f_\mathrm{rlx} T_{e0} \frac{\sum_{si} Z_{si} n_{si}(z)-n_e(z)}{n_e(z) + \sum_{si} Z_{si} n_{si}(z)},
\end{equation}
where $\phi_p^{(n)}(z)$ is the updated ambipolar potential at the new timestep, $\phi_p^{(n-1)}$ is the ambipolar potential at the previous timestep, $f_\mathrm{rlx} \approx 0.5-1.0$ is a relaxation factor added to improve numerical stability, and $n_s$ refers to the local value of density along $z$ for species $s$. If only \eqref{eq:method1} is used to calculate $\phi_p(z)$, the code will eventually become unstable. To counteract this, the potential is occasionally relaxed with a Boltzmann like response:
\begin{equation}
    \phi_p(z) = T_{e0} \ln{\left(\frac{n_e(z) + \sum_{si} Z_{si} n_{si}(z)}{n_{e0} + \sum_{si} Z_{si} n_{si0}}\right)},
\end{equation}
which has been found empirically to ensure stability over many time steps. The generation of $\phi_p(z)$ leads to off-midplane trapping due to phenomena such as sloshing ions obtained by skew injected neutral beams \citep{Kesner1980} or sloshing electrons from directed electron cyclotron heating (ECH) \citep{Baldwin1979}. Such trapping is not included in a typical bounce-averaged formulation of the Fokker-Planck equation used by CQL3D-m. To account for these trapped populations, a Maxwellian component is added to the local off-midplane distribution function with a temperature corresponding to the local potential well depth if an insufficient density can be sourced from the remapped midplane distribution function.  

The neutral beam solver Freya is unchanged from the typical version implemented in CQL3D and described in \citet{Harvey1992,Rosenberg2004}. Neutral particles are injected using realistic neutral beam geometry and tracked until they are ionized in the plasma. The particle source list from Freya is used to calculate a bounce-averaged particle source of ions and electrons with the neutral beam velocity. A bounce-averaged particle sink corresponding to NBI charge exchange is also formulated. Similarly, the computation of the quasilinear diffusion coefficients from ray data in CQL3D-m works in effectively the same way as it did in CQL3D \citep{Harvey1992} with a few adjustments to the bounce-averaging procedure to account for mirror geometry.

\begin{figure*}
  \centering
  \includegraphics[width=0.55\linewidth]{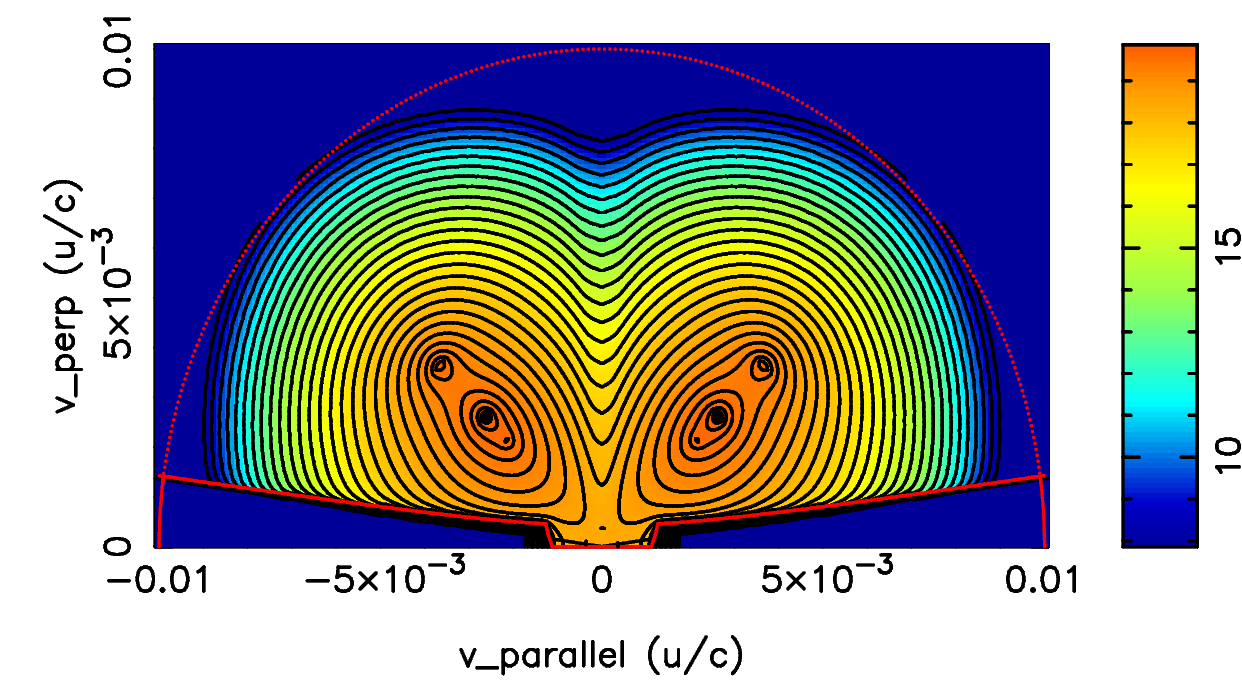}
  \includegraphics[width=0.43\linewidth]{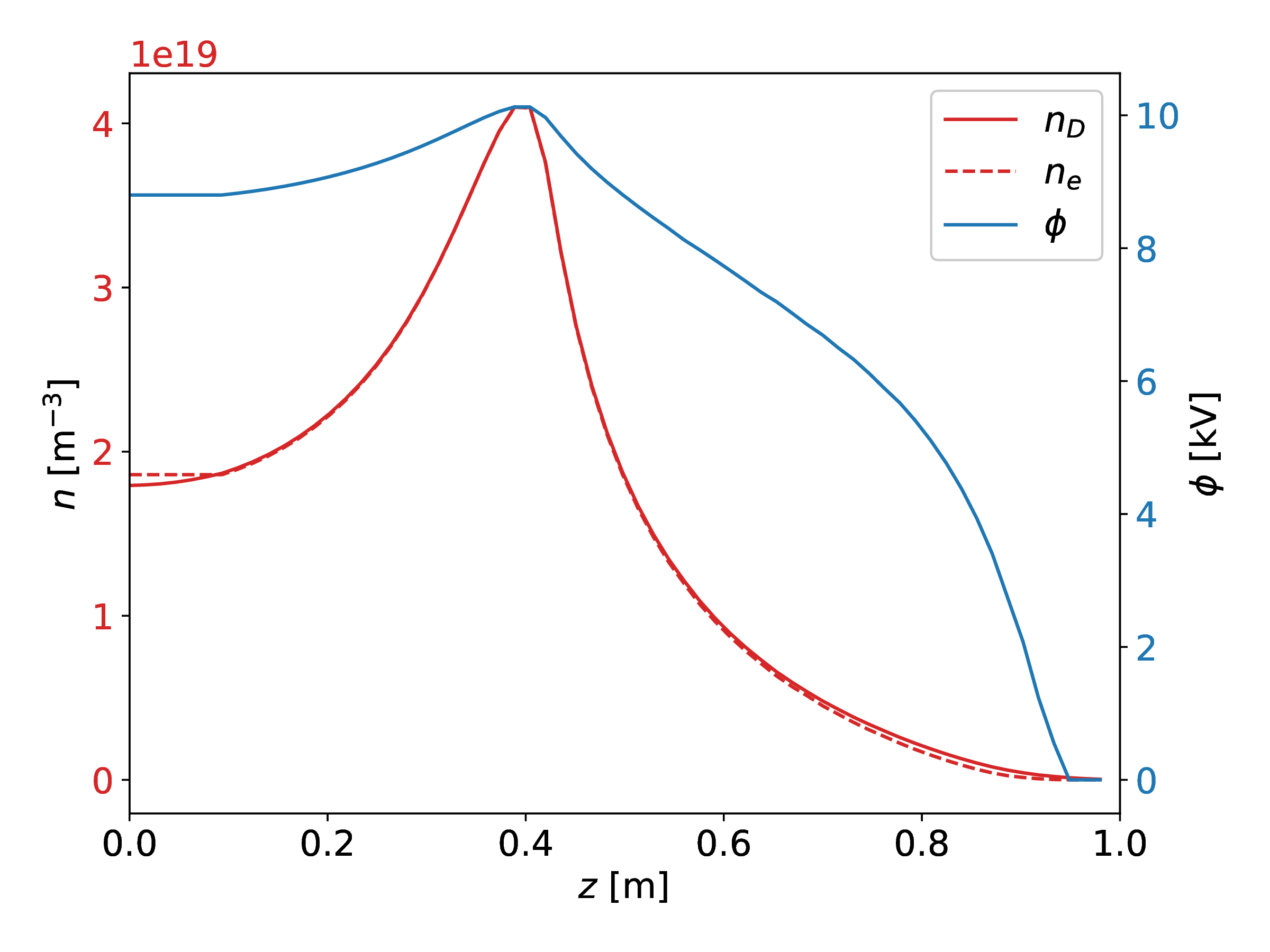}
  \caption{An example of WHAM plasma profiles and distribution functions calculated with CQL3D-m. On the left is a plot from CQL3D-m with logarithmic contours of the ion distribution in arbitrary units for the innermost $\sqrt{\psi_n} = 0.01$ normalized square root poloidal flux surface with the loss boundary shown in red. On the right are the electron (solid red) and ion (dashed red) density profiles as well as the ambipolar potential (blue) versus axial distance along the mirror $z$.}
\label{fig:cql3dm_wham}
\end{figure*}

An example of a CQL3D-m simulation of WHAM showing a sloshing ions distribution is shown in Figure~\ref{fig:cql3dm_wham}. This provides an example of many of the new features that are present in CQL3D-m. The plasma is fuelled by a $45^\circ$ NBI producing a sloshing-ions distribution function with strong density peaking about the off-midplane fast ion turning points at $B/B_0 \approx 2$. The peaking causes a corresponding off-axis potential peak to enforce quasineutrality as the electrons are still effectively Maxwellian which in turn generates a midplane centred potential well which traps lower-energy warm-ions near the midplane.

\subsection{MHD equilibrium using Pleiades}

In order to calculate the equilibrium magnetic fields of the high beta plasmas under consideration in this work, a custom fork of the Green's function based magnetic field solver code Pleiades introduced in \citet{Peterson2019} was developed. Pleiades was used in a number of previous works on mirrors \citep{Egedal2022,Endrizzi2023,Forest2024}, but the anisotropic equilibrium solver and self-consistent coupling found in this work represents a significant evolution of the code (these works included diamagnetic corrections to the magnetic equilibrium due to the plasma pressure but used a different method to solve for the corrections, and the corrections were not made self-consistently). This version of Pleiades solves for the MHD momentum balance in flux coordinates and outputs standard format \texttt{eqdsk} files containing the MHD equilibrium. The MHD momentum balance in equilibrium is defined: 
\begin{equation}\label{eq:mhd_momn}
    \boldsymbol{J} \times \boldsymbol{B} = \bnabla \bcdot \underline{\underline{\boldsymbol{P}}},
\end{equation}
where $\boldsymbol{J}$ is the current density, $\boldsymbol{B}$ is the magnetic field, and $\underline{\underline{\boldsymbol{P}}}$ is the anisotropic pressure tensor. To solve \eqref{eq:mhd_momn} using this technique, the divergence of the anisotropic pressure tensor is rewritten in component form:
\begin{equation}\label{eq:mhd_momn_int1}
    \boldsymbol{J} \times \boldsymbol{B} = \bnabla p_\perp + (p_\parallel - p_\perp)\hat{\boldsymbol{b}} \cdot \bnabla \hat{\boldsymbol{b}} + \boldsymbol{B} \bcdot \bnabla \left(\frac{p_\parallel - p_\perp}{B} \right) \hat{\boldsymbol{b}},
\end{equation}
where $\hat{\boldsymbol{b}} = \boldsymbol{B}/|B|$. Taking the cross product of $\boldsymbol{B}$ with both sides of \eqref{eq:mhd_momn_int1} after some vector algebra yields:
\begin{equation}\label{eq:mhd_momn_int2}
    \left[1 + \frac{\mu_0 (p_\perp + p_\parallel)}{B^2} \right] \boldsymbol{J}_\perp = \frac{\boldsymbol{B}}{B^2} \times \left[\bnabla p_\perp + \frac{\bnabla B}{B} (p_\parallel - p_\perp) \right].
\end{equation}
To solve the above equation in Pleiades, the correct choice of flux coordinates must be made such that derivatives of $p$ and $B$ are smooth and can be taken accurately. For an axisymmetric mirror, the most convenient choice of orthogonal coordinates for these purposes is the canonical Clebsch coordinate system ($\psi$, $\phi$, $s$) \citep{Dhaselerr}. In this coordinate system, the poloidal flux $\psi$ is the radial-like coordinate, the poloidal angle $\phi$ is the symmetry direction, and the distance along the field line $s$ is directed in the direction of the magnetic field $\hat{\boldsymbol{b}}$. Applying these coordinates to the problem dramatically simplifies \eqref{eq:mhd_momn_int2} providing an equation for the diamagnetic current response $J_{\mathrm{dia}} \equiv J_\perp \equiv J_\phi$ given the spatial profiles of kinetic $p$ and $B$:
\begin{equation}\label{eq:mhd_jdia}
J_{\mathrm{dia}} = \left[1 + \frac{\mu_0 (p_\perp + p_\parallel)}{B^2} \right]^{-1} \left(\frac{1}{B} \frac{\partial p_\perp}{\partial \psi}\nabla_\perp \psi + \frac{p_\parallel - p_\perp}{B^2} \frac{\partial B}{\partial \psi} \nabla_\perp \psi \right).
\end{equation}

With \eqref{eq:mhd_jdia}, Pleiades can be used to calculate the magnetic equilibrium by the following method: first, Pleiades is run without the plasma included; the vacuum fields and poloidal flux, $\boldsymbol{B}_{\mathrm{vac}}(R,Z)$ and $\psi_{\mathrm{vac}}(R,Z)$, produced by the current sources magnetic field coils are solved for using the Green's functions corresponding to the magnetic field coils' currents. After the vacuum fields are determined, the anisotropic pressures versus $(\psi, s)$ are obtained from integrating the local distribution functions evaluated by CQL3D-m and  mapped onto their corresponding $(\psi, s)$ in Pleiades. The $(\psi, s)$ grid is preferred here over a $(\psi, B)$ grid for the flux functions as it was found to be more numerically robust. The value of $J_{\mathrm{dia}}$ is calculated using \eqref{eq:mhd_jdia} at each location and the grid centred area $\Delta A (\psi, s)$ is multiplied by the local value of $J_{\mathrm{dia}}(\psi, s)$ to provide a current source $I_{\mathrm{dia}}(\psi, s)$. The Green's functions associated with the $I_{\mathrm{dia}}(\psi, s)$ current sources are calculated and used along with previously calculated Green's functions associated with the magnetic coils to compute updated profiles of $\psi$ and $\boldsymbol{B}$ which account for diamagnetism. The anisotropic pressure and its derivatives are then mapped to the updated $\psi$ flux surface locations and the process is repeated until the change in $\psi$ over an iteration drops below some prescribed value $\epsilon = \Delta \psi / \psi$. For $\epsilon = 10^{-6}$, this typically takes $\sim 10$ iterations. This flux advance and current redeposition scheme is very similar to the magnetic equilibrium solution scheme used in non-inductive current drive calculations with the tokamak code ACCOME \citep{Devoto1992}. As the edge pressure profile is not solved for in the current version of the RealTwin, an exponential pressure drop at the plasma edge with an $e$-folding length equal to the ion Lamor radius $\rho_i$ is used. If the value of $\epsilon$ is made small enough and a sufficiently refined ($\psi$, $s$) grid is used in the computation, the solution to the magnetic equilibrium obtained with this method will converge to the free-boundary pressure balance solution. While the Green's function approach used by Pleiades is simpler to implement than a free-boundary Grad-Shafranov solver, it is more computationally expensive as Green's functions corresponding to the diamagnetic current response must be recalculated each time $\psi$ is advanced within the code. However, as each Green's function computation is independent, the diamagnetic currents and their associated Green's functions can be computed in parallel, so a refined $(\psi, s)$ grid can be used without incurring long computational times. 

\begin{figure*}
  \centering
  \includegraphics[width=0.55\linewidth]{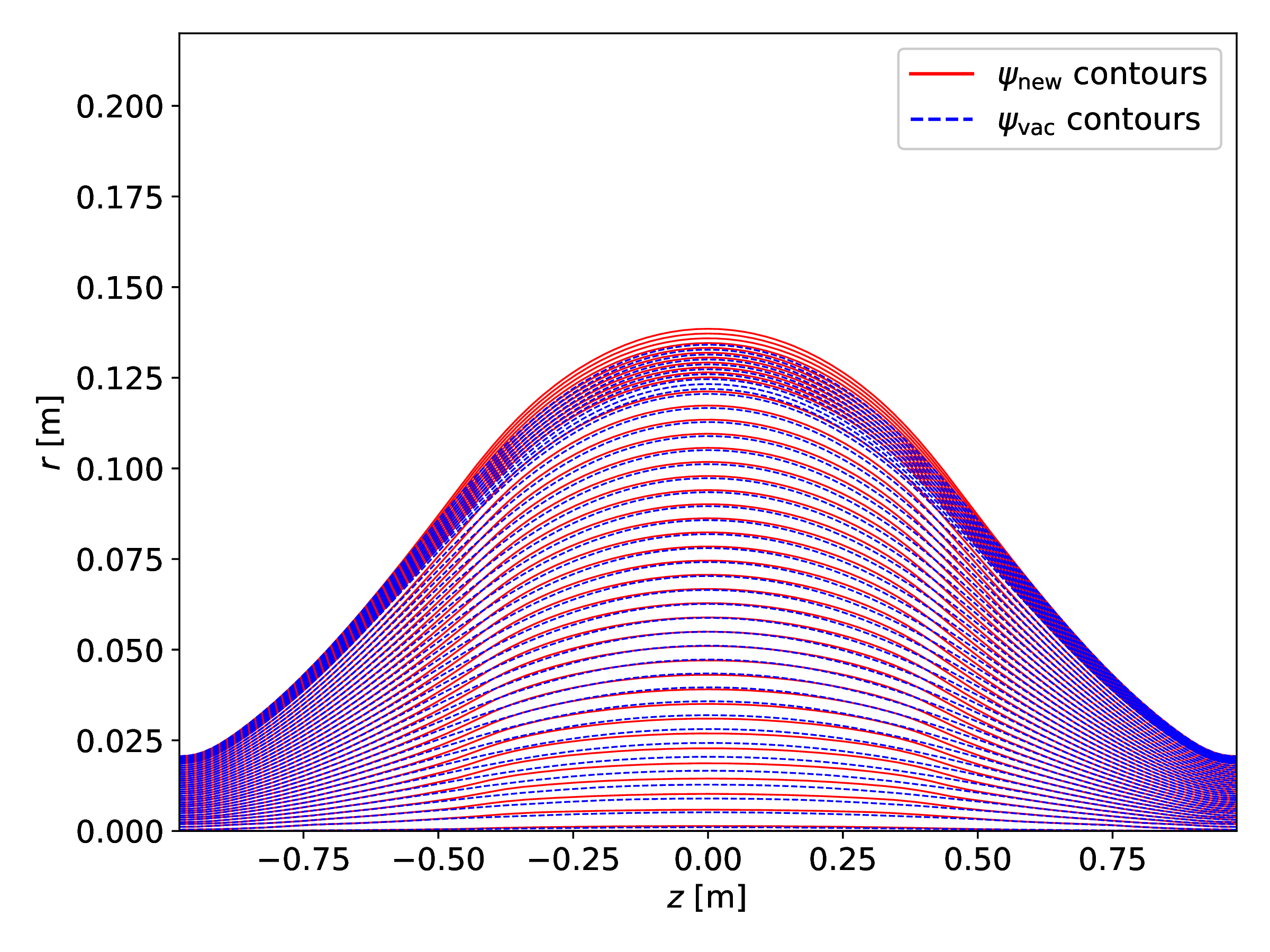}
  \includegraphics[width=0.41\linewidth]{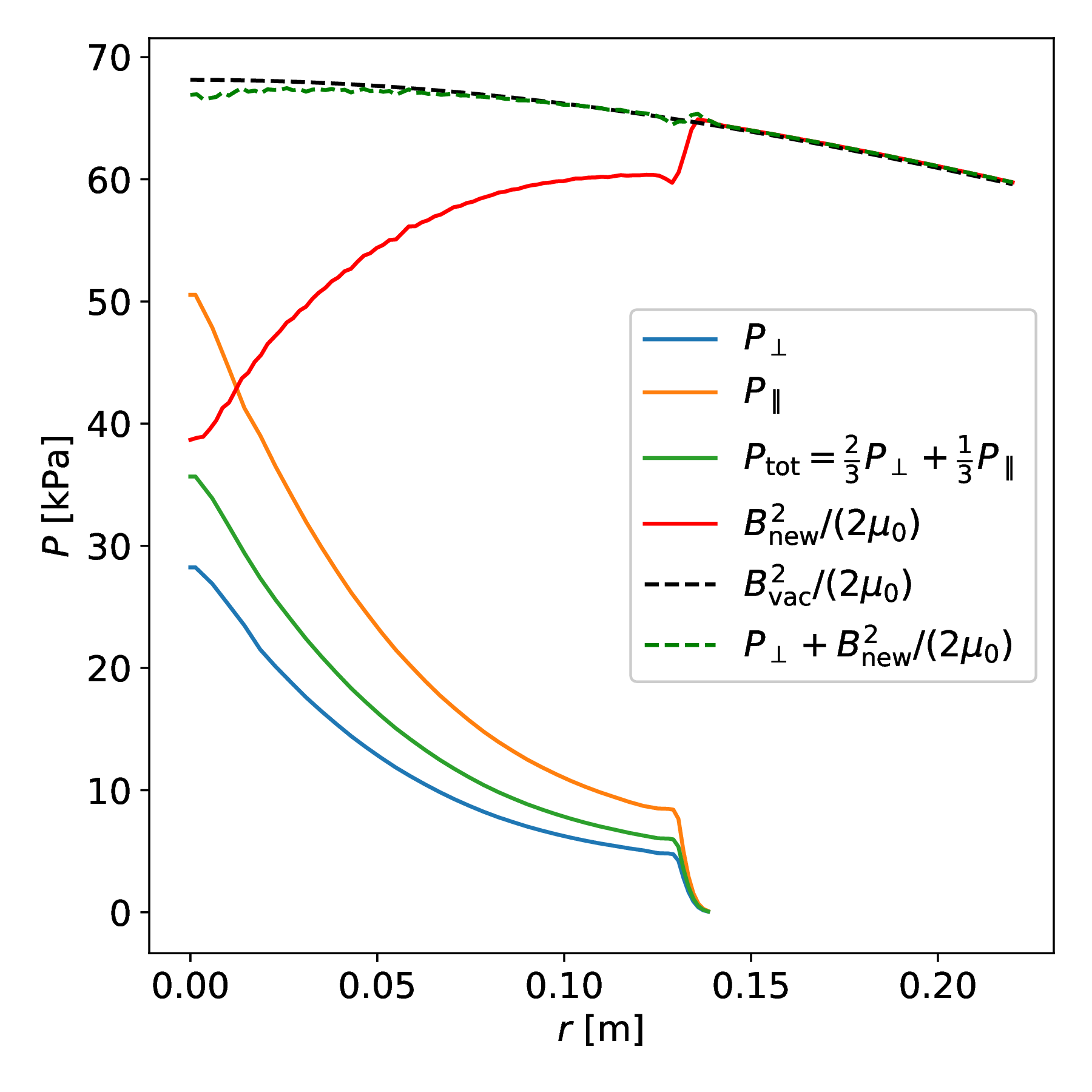}
  \caption{An example of a magnetic equilibrium for a high $\beta$ plasma in WHAM calculated with Pleiades. On the left are the flux contours for the vacuum fields $\psi_{\mathrm{vac}}$ in blue and the flux contours after diamagnetic evolution $\psi_{\mathrm{new}}$ in red. On the right are the kinetic pressure profiles and magnetic pressure profiles as well as the paraxial equilibrium condition in green which in paraxial equilibrium is equal to $B_{\mathrm{vac}}^2/2\mu_0$.}
\label{fig:pleiades_eq}
\end{figure*}

The magnetic equilibrium solver in Pleiades has been quite successful in practice and is robust to high values of $\beta \lesssim 1$. An example of a magnetic equilibrium in WHAM solved using Pleiades appears in Figure~\ref{fig:pleiades_eq} and shows the anticipated outward radial expansion of the flux surfaces consistent with a diamagnetic reduction in central magnetic field $B_0 \approx B_{\mathrm{vac}}\sqrt{1-\beta}$. The paraxial equilibrium condition from \citet{Ryutov2011}, also shown in this figure, is likewise well satisfied (though not perfectly as it is only approximate). Stability against the firehose and the mirror instabilities based on conditions \citep{Grad1967}:
\begin{equation}
    \sigma > 0
\end{equation}
\begin{equation} \label{eq:mirrorinstab}
    \frac{\partial}{\partial B}\left( \sigma B\right) > 0,
\end{equation}
 where $\sigma = 1/\mu_0 + (p_\perp - p_\parallel)/B^2$, is checked at simulation runtime. Simulations which are unstable to these MHD modes or do not converge due to $\beta>1$ are discarded (An interesting note is that Pleiades simulations unstable to the mirror mode do not readily converge. This is due to the equilibrium bifurcation phenomenon described in \citet{Kotelnikov2010} which causes the iterative equilibrium solver in Pleiades to fail).

\section{Identification of an optimized end plug with tandem mirror POPCONs}\label{sec:POPCONs}

Designing an optimized end plug requires the determination of what end plug conditions produce a tandem mirror central cell capable of high gain. To do this, a modified version of the Plasma OPeration CONtour (POPCON) technique \citep{Cordey1981,Houlberg1982} was employed. POPCONs allow the steady-state operating space of a fusion device to be mapped by enforcing a zero-dimensional power and particle balance. POPCONs have been used successfully in a number of tokamak designs, oftentimes reproducing more sophisticated transport analyses surprisingly well \citep{Sorbom2015,Creely2020,Frank2022}. This section develops a novel version of the POPCON technique for tandem mirror central cell plasmas that, given a realistic set of end plug parameters, can be used to estimate the tandem mirror central cell performance. This POPCON technique is remarkable in that it implies that central cell performance in a classical tandem is dependent on only the end plug density and the parameters of the plug mirror coils. The POPCONs will be used here to identify a set of target end plug parameters that will be demonstrated as achievable in the RealTwin simulations in the next section. 

\subsection{POPCON for tandem mirrors}
To generate tandem mirror POPCONs, the steady-state 0-D tandem mirror transport equations based on those found in \citet{Mirin1983,Hua2004} are taken to the steady state limit and all derivatives are taken to zero. However, consistent with the typical POPCON formulation, radial variation in plasma parameters (including radial variation in axial confinement) has been allowed. It is assumed that the tandem mirror is operating a classical tandem without thermal barriers \citep{Fowler1977}. Under this assumption electrons in the central cell and end plug are in thermal contact and $T_e \equiv T_{ec} = T_{ep}$. This assumption dramatically simplifies the system of equations which must be solved to make a POPCON. Under these assumptions, the primary energy and particle balance equations in the POPCON system are:
\begin{subequations}\label{eq:popcon_sys}
    \begin{align}
    &\underline{\text{Density}} \notag \\
    &2 \pi \ell_c \int_0^{a_c} \frac{n_c}{\tau_c} \, rdr = \Gamma_{\text{gas}}
    \label{eq:dens_c} \\
    &\underline{\text{Ion Energy}} \notag \\
    &2 \pi \ell_c \int_0^{a_c} \left( n_c\frac{\phi_i+T_{ic}}{\tau_c} + \mathcal{P}_{ie} -\mathcal{P}_{\text{fus},i} \right) \, rdr = P_{\text{RF},i}
    \label{eq:i_ener_c} \\
    &\underline{\text{Electron Energy}} \notag \\
    &2 \pi \ell_c \int_0^{a_c} \left( n_c\frac{\phi_e+T_{e}}{\tau_c} - \mathcal{P}_{ie} -\mathcal{P}_{\text{fus},e} + \mathcal{P}_{\mathrm{rad}} \right) \, rdr = P_{\text{RF},e},
    \label{eq:e_temp_c}
    \end{align}
\end{subequations}
where subscripts $c$ and $p$ denote central cell and end plug quantities respectively, subscripts $i$ and $e$ denote electrons and ions respectively, $n_c \equiv n_{ec} = n_{ic}$ is the density assuming quasineutrality, $T_s$ is the temperature of species $s$, $\ell_c$ is the central cell length, $a_c$ is the central cell radius, $\Gamma_{\text{gas}}$ is the gas fuelling rate, $\tau_c \approx \tau_p \approx \tau_E$ is the confinement time, $\mathcal{P}_{ie}$ is the collisional power density transfer from ions to electrons, $P_{\mathrm{RF},s}$ is the applied RF heating power to the central cell, $\mathcal{P}_{\mathrm{fus},s}$ is the fusion power density from alpha slowing down applied to species $s$, and $\mathcal{P}_{\mathrm{rad}}$ is the power density radiated by bremsstrahlung and synchrotron radiation. 

In order to create a set of POPCONs, \eqref{eq:popcon_sys} is solved for heating $P_{\mathrm{RF}}$ (with some fraction directed at ions and electrons respectively) and fuelling $\Gamma_{\mathrm{gas}}$ at a given operating point with fixed $T_{ic}$ and $n_{c}$. To solve this system, all variables must be written as a function of $T_{ic}$ and $n_c$ and fixed end plug and central cell parameters: $B_m$, $a_m$, $n_p$, $\ell_c$, $B_{0c}$ ($T_e$ may be solved for as a dependent variable). This is done by applying the following definitions; the confinement time due to classical radial and axial transport (the dominant transport modes expected in the tandem mirror) $\tau_c$ is defined:
\begin{equation}
    \tau_c = \left( \frac{1}{\tau_{\text{Past}} + \tau_f} + \frac{1}{\tau_{\rho}}\right)^{-1},
\end{equation}
where the ambipolar confinement of ions in the central cell by the end plug potential is defined \citep{Pastukhov1974,Cohen1978}:
\begin{equation}\label{eq:tau_past}
    \tau_{\text{Past}}=\frac{\sqrt{\pi}}{2}\tau_{ii}\frac{\phi_i}{T_{ic}}\exp\left(\frac{\phi_i}{T_{ic}}\right)\frac{G(R_{mc})}{1+T_i/2\phi_i-(T_{ic}/2\phi_i)^2},
\end{equation}
with $\tau_{ii}$ the central cell ion-ion pitch-angle scattering time and $\phi_i$ the ion confining potential between the central cell and the end plug resultant from their density difference:
\begin{equation}
    \phi_i = T_{e} \ln \left(\frac{n_p}{n_c}\right),
\end{equation}
and function $G(R_{mc})$ of the central cell mirror ratio $R_{mc} = B_m / (B_{0c} \sqrt{1-\beta_c})$ defined:
\begin{equation}
    G(x) = \sqrt{1+x^{-1}} \ln\left( \frac{\sqrt{1+x^{-1}} + 1}{\sqrt{1+x^{-1}} - 1} \right).
\end{equation}
The confinement due to collisional trapping of ions, which can become important for large $\ell_c$, is defined \citep{Rognlien1980}:
\begin{equation}
    \tau_f=\sqrt{\pi}R_{mc}\frac{\ell_c}{v_{\mathrm{th}ic}}\exp\left(\frac{\phi_i}{T_{ic}}\right),
\end{equation}
where $v_{\mathrm{th}ic} = \sqrt{T_{ic}/2 m_{ic}}$ is the ion thermal velocity in the central cell. The confinement time associated with classical radial transport is defined:
\begin{equation}\label{eq:tauclassical}
    \tau_\rho = 0.25 \left(\frac{a_c}{\rho_{ic}}\right)^2 \tau_{ii},
\end{equation}
where $\rho_{ic} = v_{\mathrm{th}ic}/\Omega_{ic}$ is the central cell ion Larmor radius. Returning to \eqref{eq:popcon_sys}, the electron confining potential in the plug can be defined for classical tandem operation by solving the transcendental equation \citep{Fowler1977,Hua2004}:
\begin{equation}\label{eq:phie}
    \frac{\phi_e}{T_e} \exp\left(\phi_e/T_e\right) \approx \sqrt{\frac{m_i}{m_e}}\left(\frac{T_{ic}}{T_e} \right)^{3/2}\left(\frac{\phi_i}{T_{ic}}\right)\exp(\phi_i/T_{ic}).
\end{equation}
The power density associated with ion to electron energy transfer is:
\begin{equation}
    \mathcal{P}_{ie} = -\mathcal{P}_{ei} = 3 \frac{m_e}{m_i} \frac{n_c}{\tau_{ei}}\left(T_{ic}-T_e\right),
\end{equation}
where $\tau_{ei}$ is the ion-electron collision time. The fusion power density is calculated for a D-T plasma with alphas depositing their energy on species $s$ is calculated with:
\begin{equation}\label{eq:fusheat}
    \mathcal{P}_{\text{fus},s} = f_{\alpha s}(x)\frac{E_\alpha}{4}n_c^2\langle\sigma v\rangle,
\end{equation}
using the reactivity coefficients $\langle\sigma v\rangle$ from \citet{Bosch1992} and equation for the fraction of power deposited on each species $s$ by alpha particles from the analytic solution to the slowing down equation \citep{Ott1997}:
\begin{subequations}
    \begin{equation}
        f_{\alpha e}(x) = x^{-2}\left\{x^2+\frac{1}{3}\ln\left[\frac{(x+1)^2}{x^2-x+1}\right]-\frac{2}{\sqrt{3}}\tan^{-1}\left[\frac{1}{\sqrt{3}}(2x-1)\right]-\frac{\pi}{3\sqrt{3}}\right\} 
    \end{equation}
    \begin{equation}
        f_{\alpha i}(x) = 1 - f_{\alpha e}(x),
    \end{equation}
\end{subequations}
where $x=\sqrt{E_\alpha/E_{\text{crit}}}$ and the "critical" energy where the ion and electron energy deposition by the $\alpha$ particles is balanced $E_{\text{crit}}=[3\sqrt{\pi}/(4m_i\sqrt{m_e})]^{2/3}m_\alpha T_e=59.2 T_e /\mu^{2/3}$ where $\mu = m_i/m_p$. Finally, the radiated power density is the sum of the power densities from bremsstrahlung and synchrotron radiation. The radiated power densities, using the expression for synchrotron radiation assuming an infinite straight cylinder with finite wall reflectivity, are \citep{Trubnikov1979}:
\begin{subequations}\label{eq:bremsynch}
    \begin{equation}
        \mathcal{P}_{\mathrm{rad}} = \mathcal{P}_{\mathrm{synch}} + \mathcal{P}_{\mathrm{brem}}
    \end{equation}
    \begin{equation}
        \mathcal{P}_{\mathrm{brem}} = 5.34\times10^{-32} n_e^2 \sqrt{T_e}~[\mathrm{W}\, \mathrm{m}^{-3}]
    \end{equation}
    \begin{equation}
        \mathcal{P}_{\mathrm{synch}} = 4.14\times10^{-5} \sqrt{(1-R_w)\frac{n_e}{a_c}}\left(1+2.5\frac{T_e}{m_ec^2}\right)(T_e B_{0c}\sqrt{1-\beta_c})^{2.5}~[\mathrm{W}\,\mathrm{m}^{-3}],
    \end{equation}
\end{subequations}
where $R_w$ is the wall reflectivity (taken to be 80\%, the same value used TGYRO tokamak transport simulations \citep{Candy2009}) and units of $T_e$ and $B_{0c}$ are taken to be in $\mathrm{keV}$ and $\mathrm{T}$ respectively.

Using this POPCON power-balance, it was possible to reproduce the tandem mirror operating points identified in \citet{Fowler2017} when the same assumptions were made i.e. no classical transport \eqref{eq:tauclassical}, fusion heating \eqref{eq:fusheat}, or radiation \eqref{eq:bremsynch}, and all central cell RF heating was taken to be electron directed. The $T_{ic} = 150~\mathrm{keV}$, $\beta_c \sim 1$ operating points described in \citet{Fowler2017} were found to within a small factor of error in $T_i$ (<1.25) that was determined to be associated with a discrepancy in the fusion cross sections used in their model and the Bosch-Hale cross sections used here, they considered $T_{ic} \equiv T_{ec}$ when in reality $T_{ic} \sim T_{ec}$ at these parameters, as well as the sometimes poor convergence of root-finding algorithm used to evaluate the model here at very high values of $\beta_c$. When the additional terms used here were included it was found that the operating points in \citet{Fowler2017} could not be attained. Radiation and classical transport significantly reduced performance (both terms become large at the very high $\beta \sim 1$ operating points considered there, though, it should be noted that the expression for classical transport used here is not valid when $\beta\sim1$ and \citet{Fowler2017} assumed very high wall reflectivities $R_w$).

\subsection{Identifying operating points with model end plugs}

\begin{figure*}
  \centering
  \includegraphics[width=0.45\linewidth]{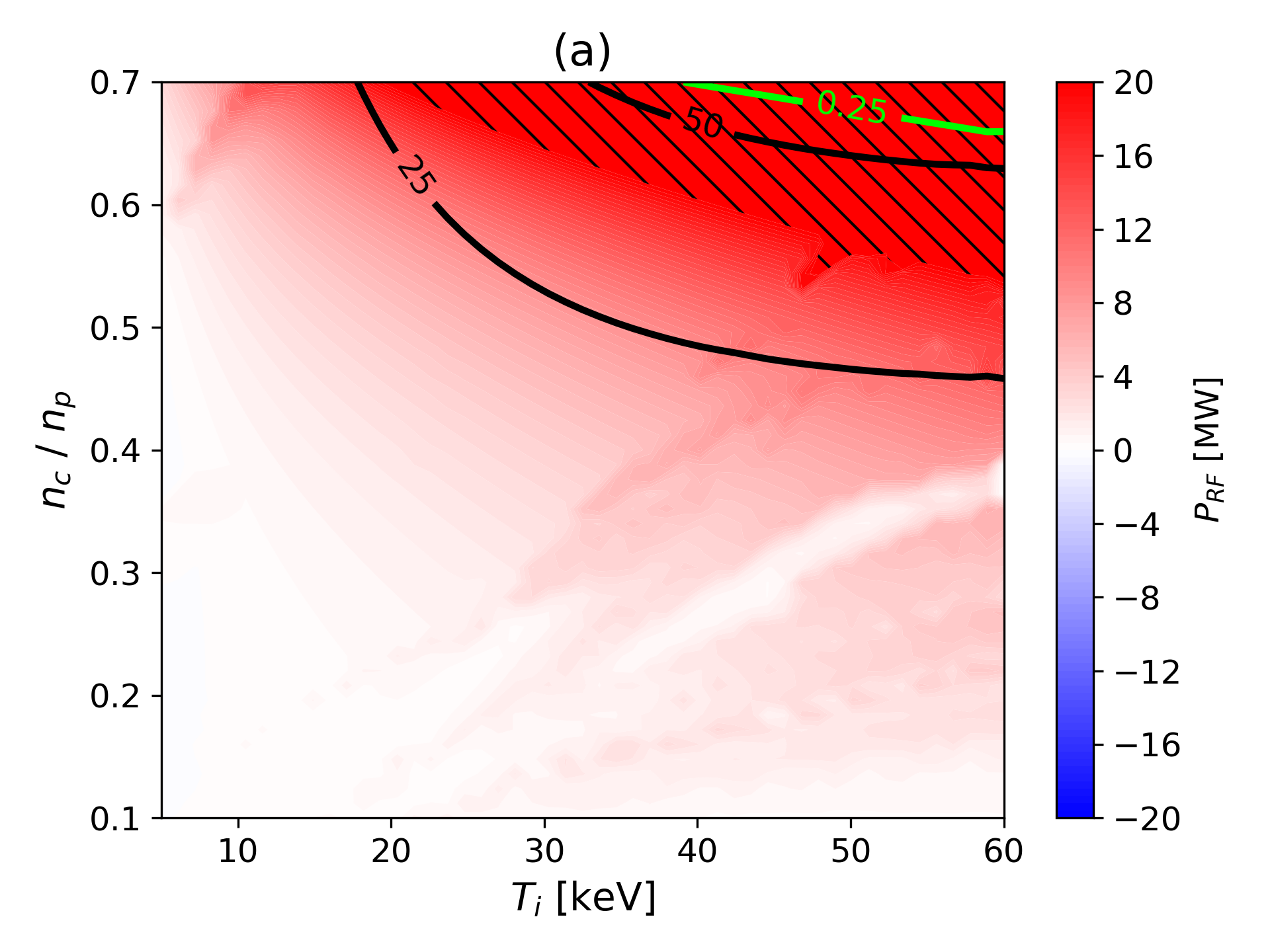}
  \includegraphics[width=0.45\linewidth]{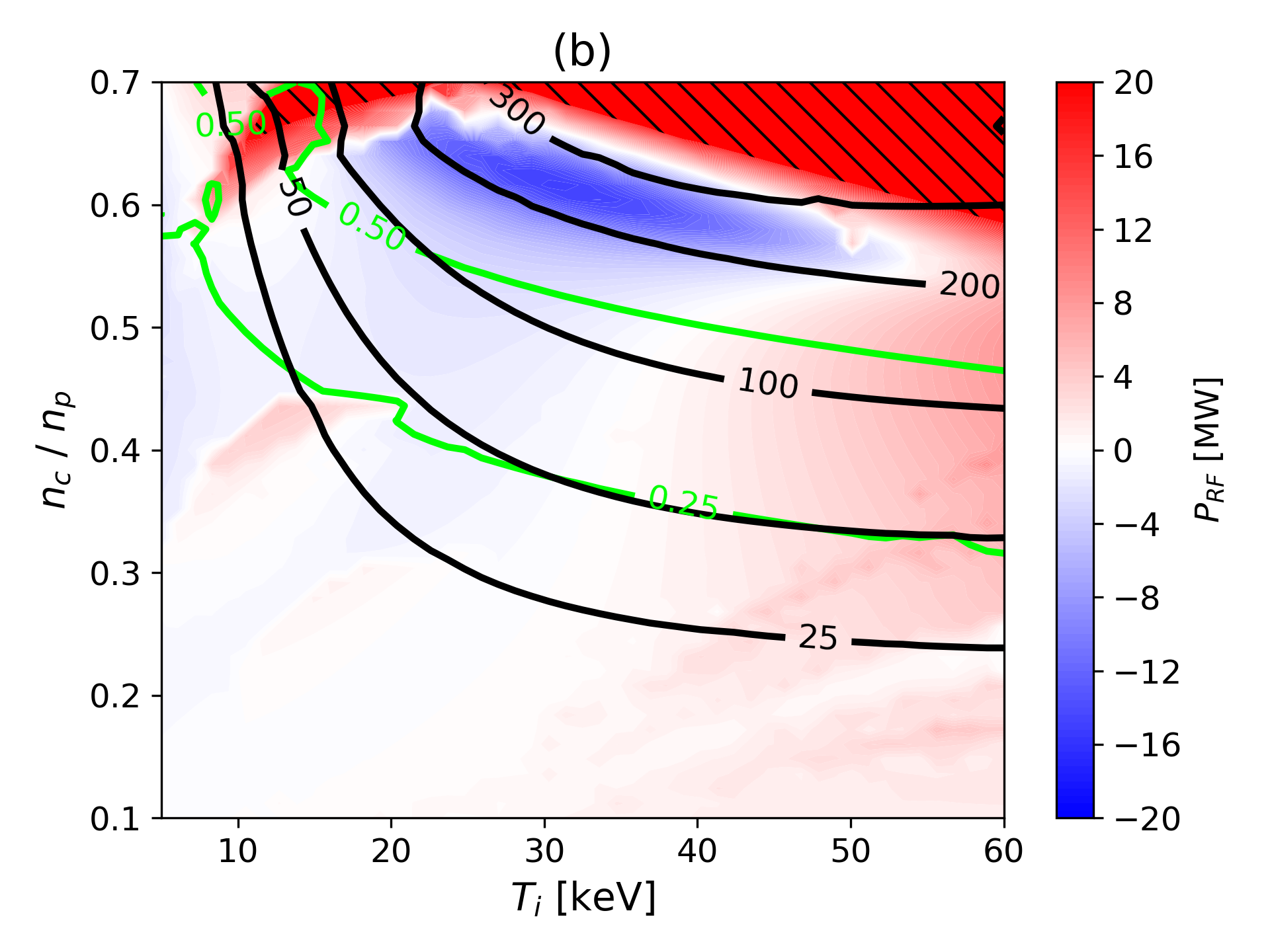}
  \includegraphics[width=0.45\linewidth]{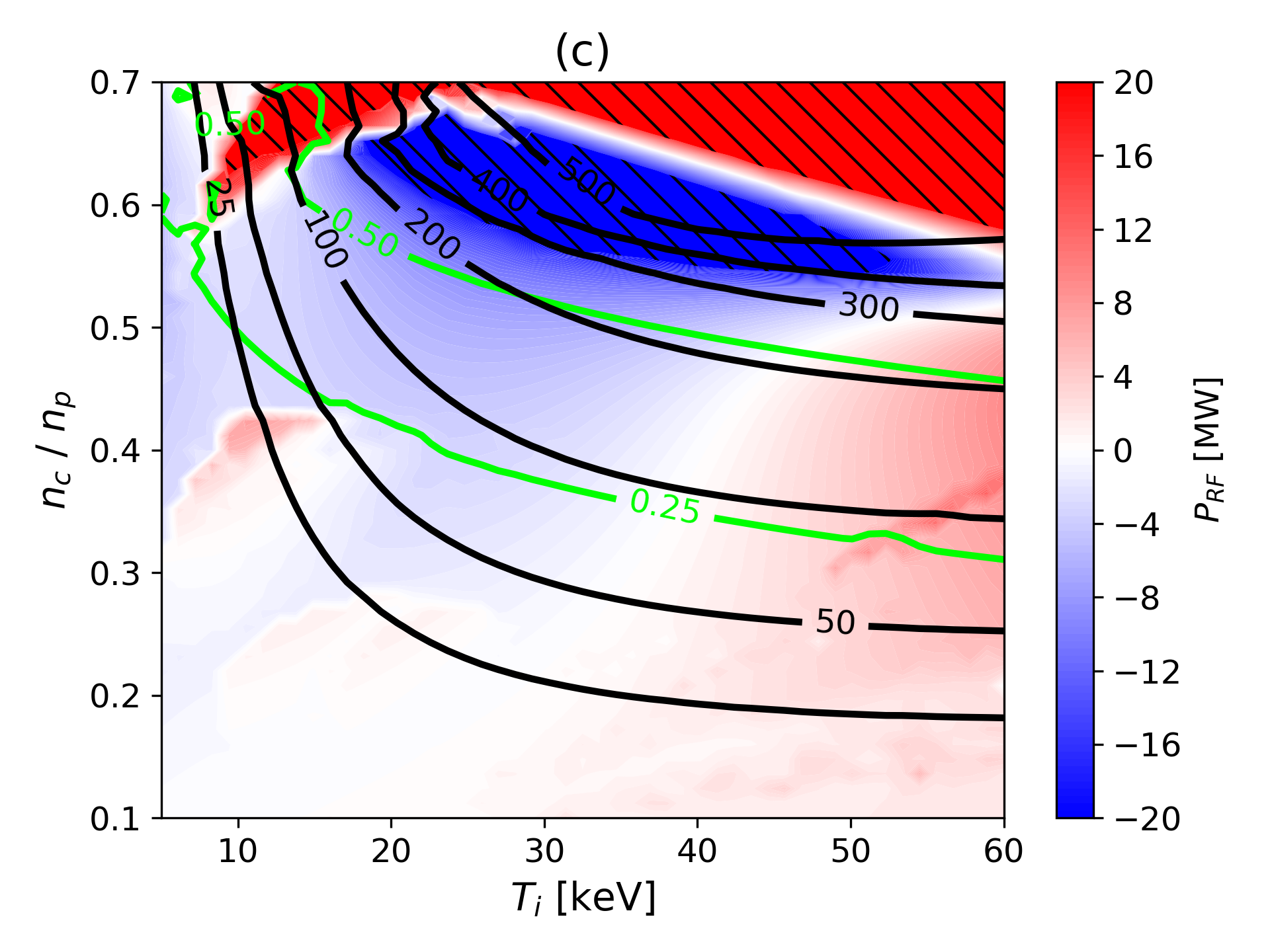}
  \includegraphics[width=0.45\linewidth]{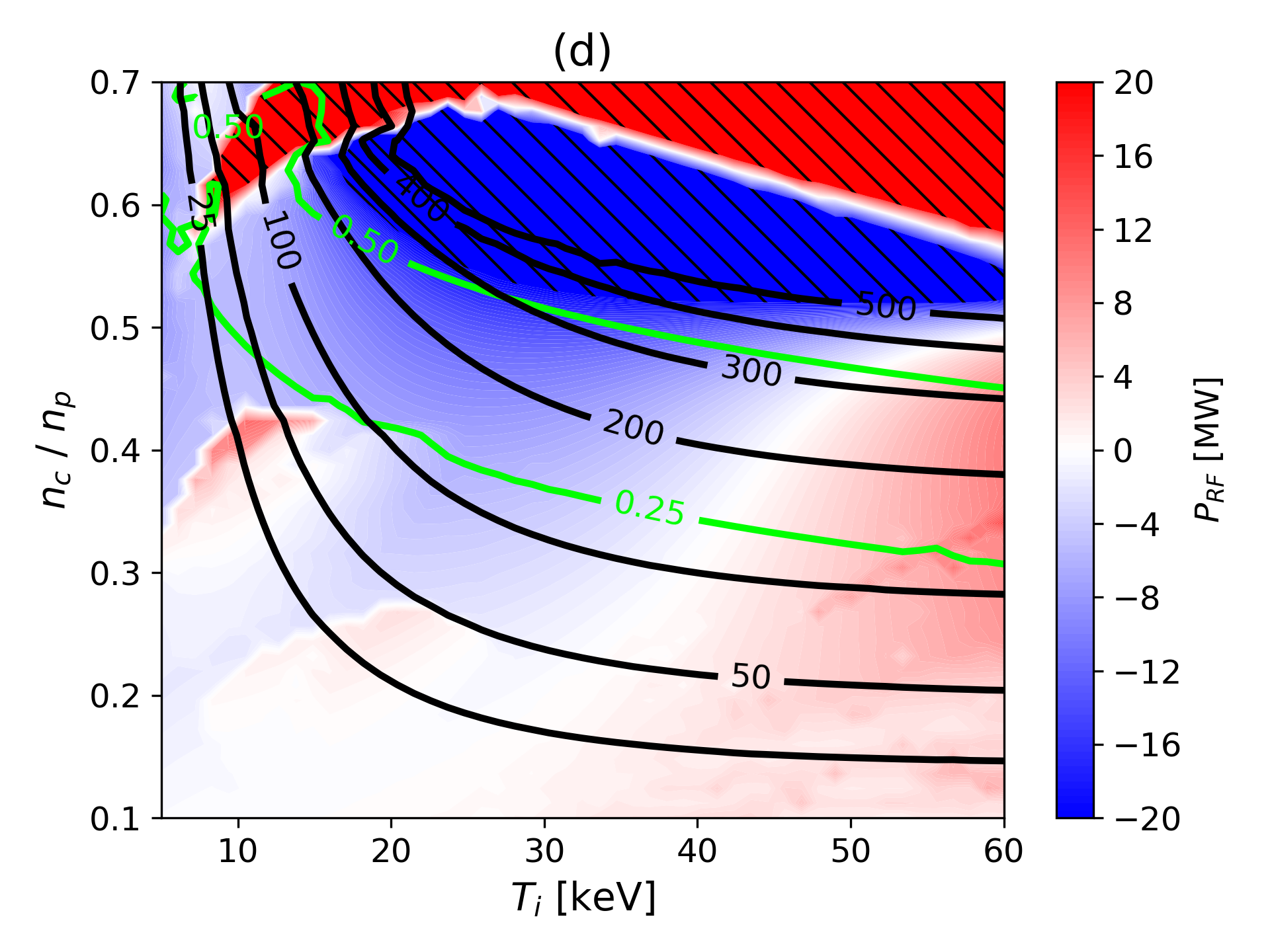}
  \caption{POPCON plots for tandem mirrors with $\ell_c = 50~\mathrm{m}$ at four different plasma radii through the mirror throat, (a) 0.1 m, (b) 0.15 m, (c) 0.2 m, and (d) 0.25 m using $n_p = 1.5\times10^{20}~\mathrm{m}^{-3}$ and $B_{0c} = 3.125~\mathrm{T}$ in all cases except for (a) where $B_{0c} = 5.0~\mathrm{T}$. Operating points are shown as a function of central cell density $n_c$ divided by end plug density $n_p$ versus central cell ion temperature $T_i$. Blue and red filled contours are of heating power to the central cell required at a given ($n_c$, $T_i$) operating point (blue regions indicate ignition). Regions which are off the scale for positive or negative and are inaccessible are denoted with red and blue hatching respectively. Black contour lines are fusion power in the central cell. Green contour lines are central cell vacuum $\beta$.}
\label{fig:popcons_ini}
\end{figure*}

With the components of the POPCON system of equations \eqref{eq:popcon_sys} defined, it is possible to solve for the required heating and fuelling at each location in ($n_c$, $T_{ic}$) space for a given set of end plug radial density profiles and engineering parameters. POPCONs with simplified end plugs were solved to establish target parameters for the end plug to be simulated with RealTwin. In these simplified end plugs, the density profile was taken to be flat, the ion directed fraction of $P_{\mathrm{RF}}$ was taken to be $80\%$. End plug parameters were taken to be fixed besides the plasma radius through the mirror throat $a_m$. The central cell field was increased until either the number of alpha gyroradii across the plasma radius $N_\alpha > 5$ defined using:
\begin{equation}
    N_\alpha = \frac{a_{0c}}{\rho_\alpha} \approx 5.24 a_m \sqrt{B_m B_{0c}}
\end{equation}
was obtained or the central cell vacuum mirror ratio reached 8 (this limit was imposed in the interest of simplifying control of central cell interchange driven by trapped particle modes based on the results in \citep{Gerver1989}). Average end plug density was varied until ignited operating points appeared. However, for illustrative purposes the POPCONs in this section all have the same density. 

The results of the POPCON analysis described here are shown in Figure~\ref{fig:popcons_ini}. POPCON analysis indicates that, for what will be shown later on are fairly reasonable end plug parameters, it is possible to create a classical tandem mirror with an ignited central cell. Ignition regions are present in all cases when the plasma radius at the mirror throat is larger than 0.15 m. These results indicate that a tandem mirror pilot plant designed with these principles should be able to satisfy the NASEM fusion pilot plant requirements at lengths under 100 m. Using simplified equation for the net electric power:
\begin{equation}
    P_{\mathrm{net,elec}} = \left(P_{\mathrm{fus},\alpha} + P_{\mathrm{NBI}} + C_{\mathrm{mult}} P_{\mathrm{fus},n} \right)\eta_{\mathrm{ele}} - \frac{P_{\mathrm{NBI}}}{\eta_{\mathrm{NBI}}}
\end{equation}
Assuming neutral beam efficiencies $\eta_{\mathrm{NBI}} = 60 \%$, an electrical conversion efficiency $\eta_{\mathrm{ele}} = 50 \%$ assuming the use of a Brayton cycle \citep{Schleicher2001}, a blanket neutron energy multiplication factor $C_{\mathrm{mult}} = 1.1$, and $P_{\mathrm{NBI}} = 30~\mathrm{MW}$ into the end plugs; it was found a tandem mirror would need to produce $157.4~\mathrm{MW}$ of power to satisfy the NASEM $50~\mathrm{MWe}$ requirement. Such an operating point is achievable with $B_m = 25~\mathrm{T}$, $a_m = 0.15~\mathrm{m}$, $\langle n \rangle_p = 1.5\times 10^{20}~\mathrm{m}^{-3}$, $\ell_c = 50~\mathrm{m}$, $\beta_c \cong 0.6$, $n_c/n_p \cong 0.55$, $T_i \cong 45~\mathrm{keV}$, and $T_e \cong 125~\mathrm{keV}$ based on this POPCON analysis. The fusion power density of this case is $\sim 3.5~\mathrm{MW m}^{-3}$ which is competitive with high-field tokamaks \citep{Frank2022}. Performance of such a system could be further enhanced by large $a_m$ or with technologies such as direct electrical conversion  \citep{Moir1973} or more efficient NBI utilizing advanced neutralization techniques and energy recovery \citep{Fumelli1989,Grisham2009,Kovari2010}. Values of $a_m < 15~\mathrm{cm}$ have difficulty obtaining good performance without very high plug densities $\langle n_p \rangle\gg 2.0\times 10^{20}~\textrm{m}^{-3}$ due to radial central cell transport and enhanced synchotron radiation. It is noteworthy that good performance is achievable here despite using generally more conservative parameters than the DT classical tandem proposed in \citet{Fowler2017} which had: $B_m = 24~\mathrm{T}$, $a_m = 0.21~\mathrm{m}$, $\langle n_p \rangle = 2.6\times 10^{20}~\mathrm{m}^{-3}$, $\ell_c = 55~\mathrm{m}$, $\beta_c \sim 1.0$, $n_c/n_p \sim 0.35$, $T_i = 150~\mathrm{keV}$, and $T_e = 150~\mathrm{keV}$.

\begin{figure*}
  \centering
  \includegraphics[width=0.55\linewidth]{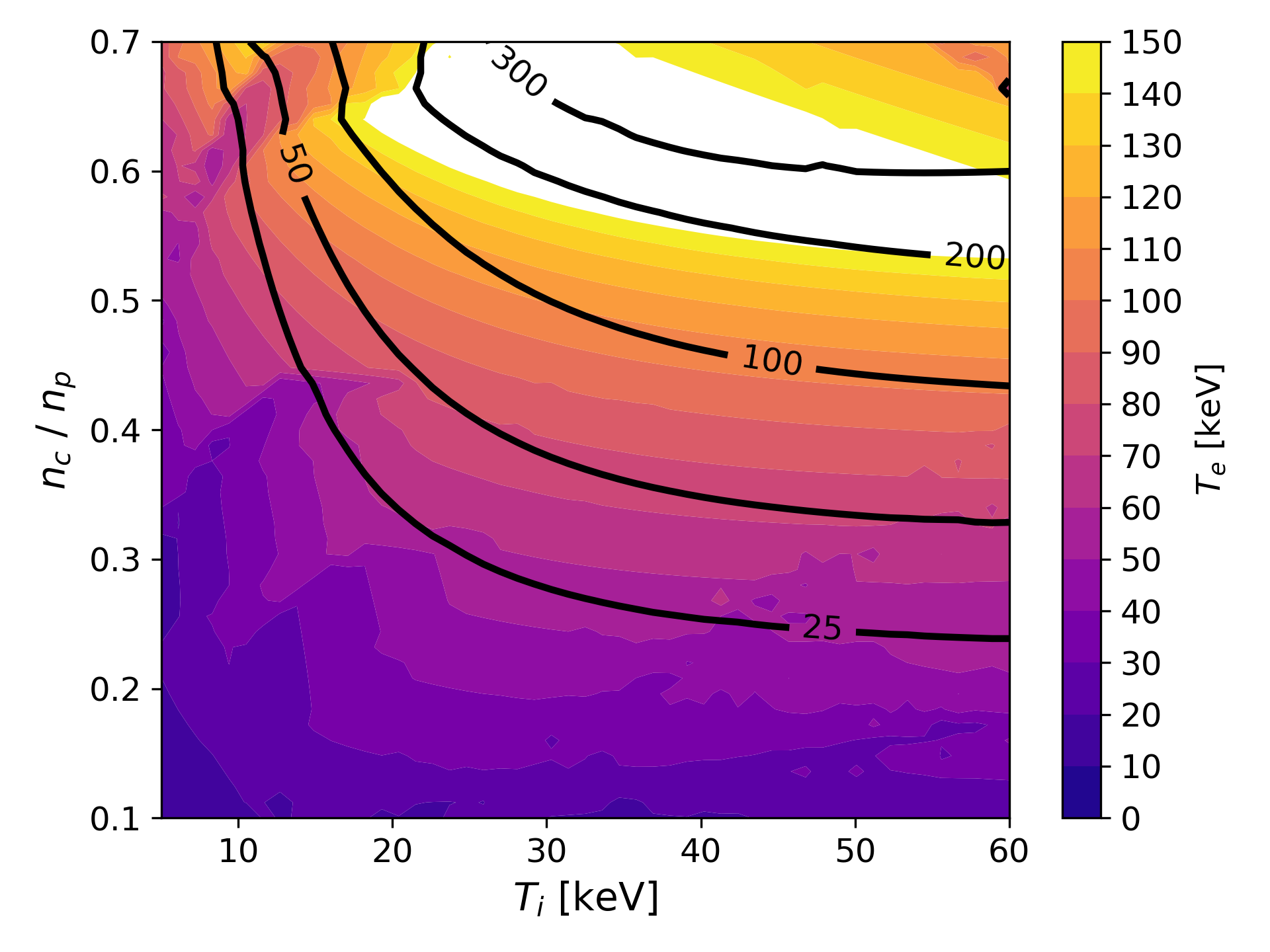}
  \caption{Contours of $T_e$ for operating points at given $\langle n \rangle$ and $E_{\mathrm{NBI}}$ satisfying the constraint equations \eqref{eq:constraints_v2}. This plot uses the $a_m = 0.15$ m POPCONs case found in Figure~\ref{fig:popcons_ini}b. Black contour lines are fusion power in the central cell in MW. }
\label{fig:popcons_Te}
\end{figure*}

This POPCONs analysis exposed a number of surprising features of fusion relevant tandem mirror operation that were not previously well known. These discoveries are in part because, outside of the analysis in \citet{Fowler2017}, axisymmetric tandem mirrors with high performance HTS magnets and high $R_m$ had never been considered. Older studies of tandem mirrors generally assumed end plug mirror ratios of order unity to account for minimum-B coil geometries and lower peak $B_m$ \citep{Fowler1977}. This analysis builds upon that in \citet{Fowler2017}, and very importantly, it includes the effect of fusion alpha particle heating on the electron temperature. The key result here is that, when high-field mirror coils are available, classical tandem central cells can and should be operated at \textit{much} higher density fractions $n_c/n_p$ than previously assumed. Much of the previous classical tandem mirror literature used $n_c/n_p = 0.05-0.35$ \citep{Dimov1976,Fowler1977,Hua2004}. The analysis here suggests this value should be higher, $n_c/n_p = 0.35-0.6$. Strong alpha heating of the electrons obtained by, increasing the central cell density, boosting confinement with large $R_m$ from axisymmetrization, and directly heating the central cell ions with RF to enhance the fusion rate, allows the central cell to reach electron temperatures $\sim 2 - 3~T_i$ without supplemental electron heating as shown in Figure~\ref{fig:popcons_Te}. This operational mode takes advantage of tandem mirror confinement \eqref{eq:tau_past} improving more rapidly with increasing temperature than it degrades with increasing the density ratio. 

In conclusion, this scoping shows that for high-field axisymmetric mirrors, classical tandem operation can reach appreciable Q and is a valid alternative to operation with thermal-barriers \citep{Baldwin1979}. These findings are in agreement with similar analyses by \citet{Pratt2006} and \citet{Fowler2017}. However, it has been found here that higher density operation than these previous analyses is possible thanks to alpha heating, somewhat relaxing the end plug performance criteria, but there are still some important caveats. The tandem mirror POPCON approach used here, much like the tokamak POPCON approach, is only able to identify operating points. It does not guarantee that an operating point it identifies is stable or achievable in actual operation. To determine this, a time dependent simulation which takes the mirror from an initial breakdown plasma to its final operating point must be performed. It is possible that, to actually reach the operational conditions proposed here, a transient heating source, such as a high-power electron cyclotron heating "sparkplug", might be needed to boost $T_e$ initially. This would drive up the end plug potentials and improve confinement enough for the central cell to generate sufficient alpha heating power to reach ignition. Furthermore, the effects of spatial gradients or turbulent transport were not included in this analysis. Generally, the spatial density gradients in tandem mirrors are predicted to be small \citep{Mirin1983,Hua2004}, and the effects of electron temperature gradient turbulence, the main form of turbulent energy transport predicted to be present in tandem mirrors, are anticipated to be more tolerable than the temperature gradient driven turbulent transport in tokamaks \citep{Hua2004,Pratt2006}. These topics will be treated in future work where gyrokinetic turbulence simulations at mirror central cell conditions will be performed to evaluate the growth rates of gradient driven instabilities. A time dependent solver based on those found in \citet{Mirin1983} and \citet{Hua2004} in order to evaluate the accessibility of these operating points under more realistic assumptions will also be implemented. In addition to these confinement issues, macrostability of the tandem system, specifically its stability against trapped particle modes and MHD, have not been addressed here and will be the topic of a future publication. 

One potentially interesting feature of the relatively high $n_c/n_p$ ratio used in this study is the question of whether outflow from the central cell may have a stabilizing affect on kinetic modes in the end plugs. The flux from the central cell through the end plug midplane can be determined by particle balance of the outflow along the flux tube:
\begin{equation}
    \Gamma_{c,\mathrm{out}} = \frac{n_c \ell_c}{2 \tau_c} \left(\frac{R_{mc}}{R_{mp}} \right)  
\end{equation}
If this outflow flux becomes large enough, it has been shown to be stabilizing to DCLC \citep{Drake1981}, but large amounts of central cell outflow could also reduce an end plug's classical performance by increasing the ion-ion collisional scattering rate. Taking the estimate for the outflow required to stabilize the plug from \citet{Correll1980}:
\begin{equation}
    \Gamma_{p, \mathrm{stab}} > 1.2 \times 10^5 \frac{n_p \phi_e^{3/2}}{(R_{mp}\sin^2\theta_{\mathrm{NBI}}-1) \sqrt{\mu} E_{ip}} \left(\frac{a_m\sqrt{R_{mp}}}{\rho_{i0}}\right)^{-4/3}~[\mathrm{m}^{-2} \mathrm{s}^{-1}],   
\end{equation}
where $\theta_{\mathrm{NBI}}$ is the angle of the NBI in the plug, $\mu$ is the average ion mass in the plug in AMU, $\phi_e$ is the plug potential in kV, $E_{\mathrm{NBI}}$ is the plug NBI energy in keV (all other quantities are assumed to be in MKS), and $\rho_{i0}$ is ion Larmor radius at the midplane of the plug. Using these two equations, the minimum $n_c/n_p$ density ratio at which DCLC stabilization in the plug can be obtained: 
\begin{equation}
    \frac{n_c}{n_p} > 2.4 \times 10^5 \frac{\tau_c \phi_e^{3/2}}{\ell_c (R_{mp}\sin^2\theta_{\mathrm{NBI}}-1) \sqrt{\mu} E_{ip}} \left(\frac{R_{mp}}{R_{mc}} \right) \left(\frac{a_m\sqrt{R_{mp}}}{\rho_{i0}}\right)^{-4/3}.
\end{equation}
For tandem mirror parameters similar to the ignited operating points in Figure~\ref{fig:popcons_ini}b: $\tau_c = 5~\mathrm{s}$, $\phi_e = 400~\mathrm{keV}$, $\ell_c = 50~\mathrm{m}$, $R_{mp} = 8$, $R_{mc} = 13.3$ (accounting for finite $\beta \sim 0.6$ in both mirror ratios), $\theta_{\mathrm{NBI}} = 45^\circ$, $E_i = 500~\mathrm{keV}$, $\mu = 2.5~\mathrm{AMU}$, and $a_m \sqrt{R_{mp}}/\rho_{i0} = 25$, it is calculated that $n_c/n_p > 1000$ indicating DCLC likely cannot be stabilized with central cell outflow in the tandem mirrors under consideration here. As $\Gamma_{c,\mathrm{out}}$ is insufficient for DCLC stabilization, DCLC stability in Section~\ref{sec:optimization} will be ensured by the suppression of the diamagnetic drift wave.

Finally, there is the matter of alpha particle removal from the central cell in steady-state operation. Alpha particles are well confined in the central cell by the end plug's ambipolar potential. Without a method to deconfine them, they will accumulate. This will need to be addressed with a technique such as drift pumping, which removes thermalized alphas by applying a radiofrequency perturbation to the background magnetic field that causes them to drift out of the plasma after enough central cell transits \citep{Dimov1976,Logan1983}. 

\section{Simulation of an end plug with optimized equilibrium, heating, and transport using RealTwin}\label{sec:optimization}

The POPCON technique outlined in the previous section has defined what parameters the end plug must achieve in order to obtain a good central cell: the end plug mirror coils must have the highest field and bore possible, and an end plug must achieve the highest plasma density possible for the fewest watts of input power. In this section an end plug's equilibrium transport will be optimized, first some assumptions will be used to simplify the problem and 0D physics and engineering constraints will be imposed on the end plug. The 0D constraints will be used to identify an end plug operating region. Then, simulations with the RealTwin will be performed to provide a coarse optimization of a simple mirror based on the 0D constraints  to determine if the end plug conditions predicted by the 0D scoping can be obtained. Next, using the simple mirror density $\langle n_p \rangle$ parameters obtained by RealTwin, a new set of tandem POPCONs will be calculated. The range of tandem mirror ambipolar potentials and temperatures obtained from the POPCONs will be used to initialize a secondary set of RealTwin simulations with a fixed temperature electron backgrounds and an ambipolar potentials that correspond to the values obtained using POPCONs. These simulations will be iterated with the POPCONs until the end plug RealTwin simulation conditions and the POPCON conditions are self-consistent to provide a final equilibrium transport optimized end plug.

\subsection{Assumptions and 0D constraints}\label{sec:assumptions} 

When performing simulations with the RealTwin model, several assumptions about end plug operation were made due to the limitations of present mirror modelling tools (a discussion of ways in which these capability gaps may be closed in the future appears in Section~\ref{sec:conclusion}). In addition, a number of constraints related to engineering feasibility and plasma stability were used to do significant optimization work prior to even running the RealTwin.

Beginning with the modelling assumptions and fixed parameters; two large physics assumptions were made in this study to overcome the limitations of present modelling tools. First, it was assumed that the end plug, for the purposes of equilibrium and transport modelling, could be treated as a standalone simple mirror with certain correction factors. The justification of the first assumption lies in the fact that the ion populations in the plug and the central cell are to lowest order "detached". Ions confined in the central cell do not enter the plugs and ions confined in the plugs do not enter the central cell. Similarly ions that are not confined in the plug are also not confined in the central cell and vice-versa. This property of tandem mirrors ends up being useful for reasons that will be discussed momentarily. In fusion-relevant conditions, confinement is good in both the plugs and the central cell and the unconfined population of ions flowing through either region is small, acting as a (relatively) low density stream which passes outwards until it reaches the expander region \citep{Cohen1980}. However, in a classical tandem mirror the electron population \textit{is} able to transit between the plug and central cell, and alpha particle heating from the central cell is sufficient to enable us to reach high electron temperatures. In fact, the predicted electron temperatures in a tandem are \textit{higher} than they otherwise would be in a standalone end plug due to alpha heating of electrons in the central cell. The initial simulations in Section~\ref{sec:simp_mirror} of the standalone simple mirror end plug utilize a dynamic electron temperature. However, later simulations in Section~\ref{sec:selfconsist} use a fixed values of $T_e$ and ambipolar potential $\phi_e$ corresponding to the values computed in the tandem POPCON.

The second important physics assumption is that MHD \citep{Rosenbluth1957} and trapped particle mode \citep{Berk1986a,Berk1986b} were not considered in detail. While kinetic stability to the DCLC instability and MHD stability to the mirror and firehose instabilities are considered here, stability against flute modes and rigid $m=1$ displacements driven by MHD and trapped particle modes have not been assessed. While it is acknowledged that stability to these modes is vital to developing an operating mirror fusion device, in order to effectively assess stability and develop working stabilization actuators, self-consistent mirror equilibrium parameters, like those calculated here, must be obtained first. This work encompasses that important first step. The IPS driven RealTwin simulation framework built here was constructed for the express purpose of coupling transport, heating, and equilibrium simulations self-consistently to stability simulations. An early demonstration of this capability, in which CQL3D-m/Pleiades simulations are coupled to hybrid Vector Particle-In-Cell (hybrid-VPIC) simulations in order to assess extended MHD and kinetic stability. This is presented in this article's companion article \citet{Tran2024}. Realta Fusion plans to continue to investigate kinetic and MHD stability utilizing a full complement of simulation tools coupled to IPS driven simulations including: linear MHD (FLORA) \citep{Cohen1986}, nonlinear MHD (NIMROD) \citep{Sovinec2004}, particle-in-cell methods (VPIC, WARP-X) \citep{Bird2021,Fedeli2022,Le2023}, as well as gyrokinetic and multifluid codes (Gkeyll) \citep{Hakim2008,Francisquez2023}. These simulation tools can also account for extended effects: such as charge separation's impact on trapped particle modes \citep{Kesner1983}, FLR effects' impact on MHD stability \citep{Ryutov2011}, and sloshing ions as well as neutral particle fuelling's impact on DCLC \citep{Baldwin1977,Tran2024}. As stability actuators and these effects can interact in ways which are nonlinear and difficult to predict analytically, high-fidelity simulation tools and self-consistent simulations are vital to developing a complete picture of stability.

This simulation model will eventually allow a systematic evaluation of axisymmetric mirror stability to trapped particle modes, MHD, and kinetic modes, using realistic equilibrium parameters to be performed under the influence of stability actuators such as: end-ring induced shear flows \citep{Sakai1993,Beklemishev2010,Bagryansky2011}, localized electron cyclotron heating induced shear flows \citep{Cho2005,Cho2006}, active feedback control \citep{Lieberman1977,Kang1988}, high-$\beta$ \citep{Berk1987} field ripple in the central cell in combination with conformal conducting walls \citep{Li1987}, kinetic stabilizers \citep{Post2001}, and line-tying using cold plasma mantels \citep{Fornaca1979,Molvik1984}. In addition to MHD modes, there is interest in the impact of these actuators on trapped particle modes in the tandem mirror. Kinetic stabilizers \citep{Post2001} have been shown to be potentially ineffective at stabilizing trapped particle modes \citep{Berk2011,Fowler2017}. However, as trapped particle modes in tandem mirrors cause fluid-like interchange which has similar growth rates and is phenomenologically similar to MHD, it may be possible to stabilize them with the same feedback control, line tying, and shear flow actuators as MHD \citep{Kesner2018}. Another potential avenue for trapped particle mode stabilization is the difference in the electron and ion bounce points in the central cell which can have a stabilizing effect on trapped particle modes in classical tandem mirrors like those here \citep{Kesner1983}.

As mentioned earlier, the detached nature of the end plug and central cell ion distributions can be used to improve performance. In a tandem, \textit{the end plug ion composition need not be the same as the central cell} as confined ion populations in the end plug and central cell are not exchanged \citep{Fowler1977,Cohen1980}. This presents a unique opportunity to enhance end plug physics performance and simplify end plug engineering. To this end, end plugs fueled with only tritium are used in conjunction with a deuterium and tritium fueled central cell. This combination maximizes end plug performance and reduces the nuclear engineering challenges presented by the end plug. The physical argument for this operational mode is as follows: ion confinement in the end plug is dominated by two processes: ion-ion scattering and electron drag. The particle confinement times associated with these two processes are \citep{Logan1980}: 
\begin{subequations}\label{eq:logan_scaling}
    \begin{align}
        \tau_s &= 2.76\times10^{18}\mu^{1/2} E_p^{3/2} \frac{\log_{10}(R_m \sin^2\theta_{\text{NBI}})}{Z^4 n_i\ln\Lambda_{ii}} \\
        \tau_d &= 1\times10^{14}\mu T_{e,p}^{3/2}\frac{\ln(E_{\mathrm{NBI}} / \langle E_L \rangle)}{Z^2 n_e\ln\Lambda_{ei}} \\
        \tau_p &\approx \left(1/\tau_s + 1/\tau_d \right)^{-1},
    \end{align}
\end{subequations}
where the average ion loss energy $E_L$ can be defined:
\begin{equation}
    E_L \approx \frac{1}{1+(\tau_s/\tau_d)}\left[E_{\mathrm{NBI}}+E_h\left(\frac{\tau_s}{\tau_d}\right)\right],
    \label{eqn:E_L}
\end{equation}
with ambipolar hole loss energy:
\begin{equation}\label{eq:ambi_loss_e}
    E_h =\frac{\phi_e}{R_m\sin^2\theta_{\text{NBI}}-1}.
\end{equation}
At typical tandem operational parameters, confinement is assumed to be electron drag dominated, $\tau_d \ll \tau_s$ \citep{Logan1980,Mirin1983,Hua2004,Fowler2017}. As the end plug ion population composition need not be the same as the central cell, performance can be enhanced by choosing ions with large $\mu/Z^2$. In addition, choosing an ion mix to suppress fusion neutron rates in the end plug can substantially simplify engineering by reducing radiation shielding requirements for components such as magnets and NBI. Suppressing fusion rates in the end plugs is acceptable as the total fusion power produced by the end plug is a small component of the overall fusion power in a tandem, since the end plug volume is much less than central cell volume, $V_p \ll V_c$. If taken to its logical conclusion, this argument suggests that a tritium (T) only end plug is optimal with respect to classical transport and neutron damage. Tritium has the highest $\mu/Z^2$ ratio of any standard fusion fuel ion, and the T-T fusion cross-section is nearly an order of magnitude lower than the D-D fusion cross section, and T-T fusion produces neutrons with energies of $\sim 3~\mathrm{MeV}$ \citep{Brown2018}. Thus, a T-only end plug provides a factor of $\sim 1.2$ increase in confinement performance and a many-order of magnitude reduction in the neutron rate as well as a significant reduction in neutron energy relative to a D-T end plug (T fuelled plugs will have inferior microstability to D or H plugs as the T Larmor radius is larger, making it easier to excite diamagnetic drift waves. However, it will be shown that it should be possible to avoid diamagnetic drift waves with large $a/\rho_i$ even with T plugs). 

Another assumption is the use of only negative-ion neutral beams for end plug heating in steady-state (a discussion about the limitations associated with high-harmonic RF heating coupled with positive-ion NBI, similar to what is planned in WHAM, and the reason it is not used in this work is discussed in Appendix~\ref{sec:rf_limits}). The use of ECH has also been limited to achieving breakdown (assuming a scheme like that found in \citet{Yakovlev2016}). Presently, the use of any significant quantity of ECH in CQL3D-m causes the potential profile to invert. This behaviour is actually promising as it indicates that it may be possible to generate thermal barriers \citep{Baldwin1979}, but at the moment it causes significant problems with the bounce-averaging and ambipolar potential calculations in CQL3D-m. Thus, further investigation of ECH is left for future work. Neutral transport into the tandem mirror plasma has also been neglected. This effect is anticipated to be of some importance in small relatively low density mirrors such as WHAM, where neutrals recycled from the edge can enter the plasma and charge exchange with well-confined beam ions causing them to be lost and replacing them with a promptly lost cold ion. However, fusion-grade plasmas like those under consideration here are opaque to edge neutrals, and the effect of these neutrals on transport can be ignored. The final assumption made here is that transport in the end plug will remain classical during operation. To this end, it has been insured that the DCLC stability condition is well satisfied in the plasmas optimized here.

\begin{figure*}
  \centering
  \includegraphics[width=0.6\linewidth]{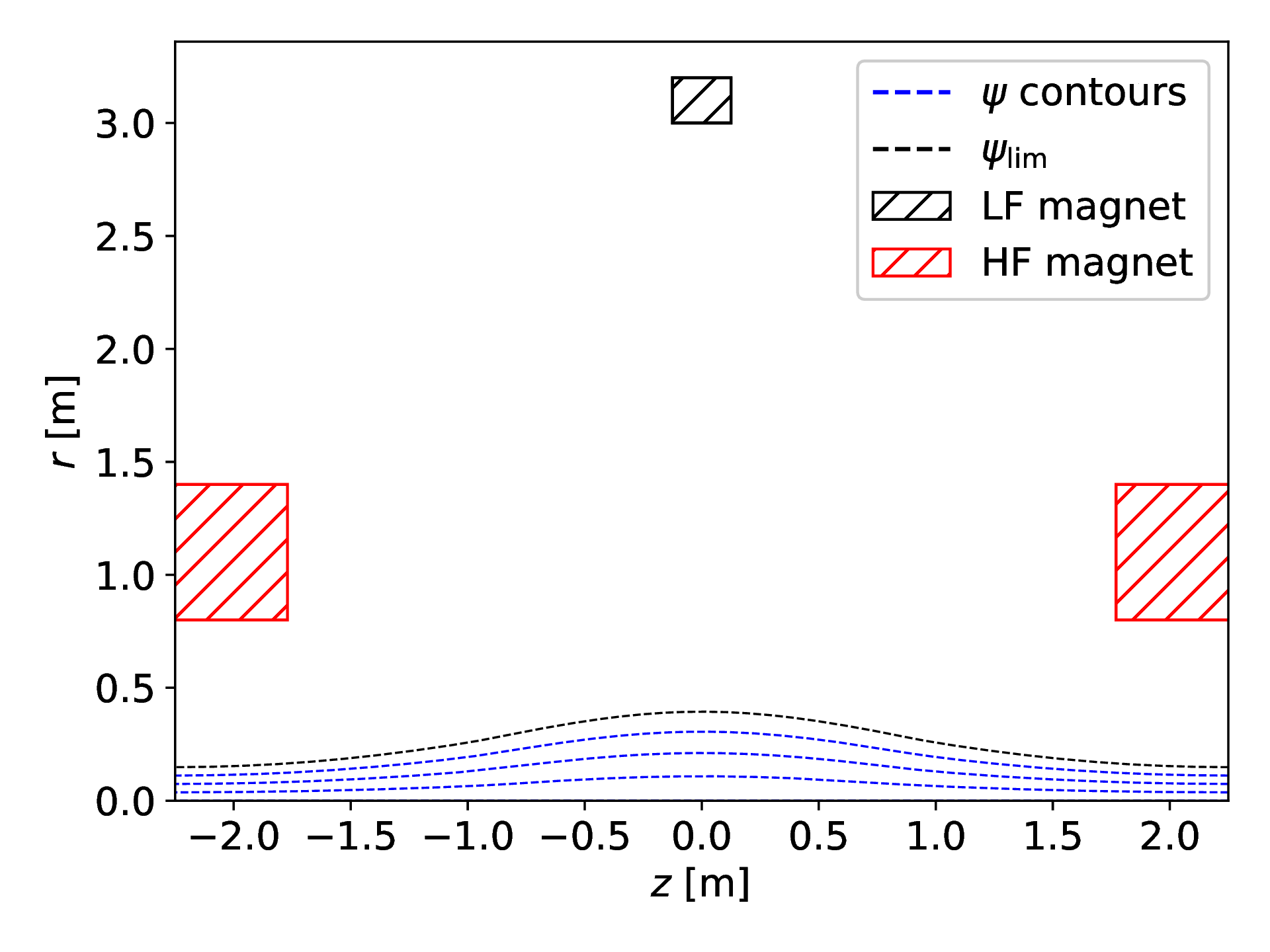}
  \caption{An example of the magnetic configuration used in the RealTwin end plug simulations. The high-field (HF) mirror coils are shown in red and the low-field (LF) central coil is shown in black. Contours of poloidal flux are shown in blue and the limiting surface is shown with a black dashed line.}
\label{fig:magnets}
\end{figure*}

In addition to physics assumptions, some engineering constraints were imposed on the mirror design here. First, the field of the two large bore high-field mirror coils placed at either end of the end plug (denoted HF) was fixed at $25~\mathrm{T}$. Feasibility studies of HTS coils at these fields have found that such designs should be viable at bores sizes appropriate for fusion applications, $\sim1$ m warm bore diameter. A single low-field coil (denoted LF) was placed at the centre of the end plug to give control of the central field. The magnetic configuration for an example case is shown in Figure~\ref{fig:magnets}. The minimum length $\ell$ (where subscript $p$ on end plug quantities in this section has been dropped for simplicity), defined as the mirror throat to mirror throat distance. This limit was chosen not by the finite Larmor radius (FLR) stabilization condition for $m \ge 2$ modes, but rather by engineering constraints. The substantial size of the neutral curvature central cell in addition to the good curvature expanders is expected to provide significant stabilization against high $m$ modes \citep{Roberts1962,Rosenbluth1962,Rosenbluth1965,Pearlstein1966,Anikeev1992,Ryutov2011,Fowler2017}. Active stabilization schemes are anticipated to provide for remaining MHD stabilization needs. Here, $\ell \ge 4.5~\mathrm{m}$ has been chosen as with values of $\ell < 4.5~\mathrm{m}$ require large reverse polarity currents in the LF coil to cancel out the HF field. Additionally, very short end plugs will almost certainly impede NBI port access and require the use of inefficient NBI launch angles that limit beam path length, causing reduced NBI absorption. The optimal NBI configuration for maximizing beam path length, and maintaining good confinement (noting there is an effective mirror ratio reduction associated with NBI launch angles smaller than $90^\circ$ relative to the magnetic axis) occurs at, or very close to, a neutral beam with $\theta_{\mathrm{NBI}} = 45^\circ$ targeted at the midplane. To maintain an effective $45^\circ$ ion velocity-space angle for different beam launch angles, the beam's target location must be moved to different values of $B/B_0$. The reduction factor in beam path length incurred by this move is in a simple mirror is $\sqrt{(1+\cot^2 \theta_\mathrm{NBI})/4}/\sin\theta_{\mathrm{NBI}}$. To simplify this study, only the near optimal case of $\theta_{\mathrm{NBI}} = 45^\circ$ with the beam directed at the midplane was examined in detail (the NBI here used a $0.4~\mathrm{m}$ by $0.8~\mathrm{m}$ source with no beam divergence as a placeholder until more realistic NBI specifications are known). In later work, with better defined, beam, magnet, and radiation shielding parameters, this angle may be slightly adjusted to prevent beam-line interference with components.

\begin{figure*}
  \centering
  \includegraphics[width=0.7\linewidth]{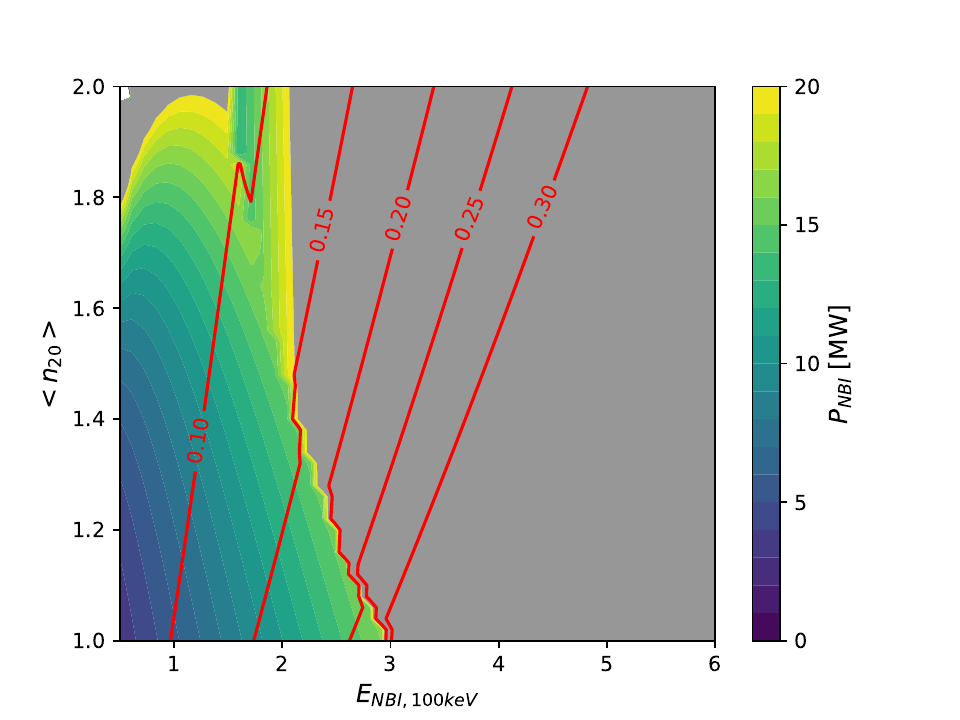}
  \caption{Contours of $a_m$ (red lines) and $P_{\mathrm{NBI}}$ (filled contours) for operating points at given $\langle n \rangle$ and $E_{\mathrm{NBI}}$ satisfying the constraint equations \eqref{eq:constraints_v2}. The grey region represents operation regimes requiring greater than 20 MW of NBI power to access.}
\label{fig:constraints}
\end{figure*}

After fixing some engineering parameters, a basic physics constraint based optimization was used to identify a parameter space which seemed likely to produce attractive results. To do this, the primary parameters of an negative-ion NBI driven plug: $a_m$, $\ell$, $B_m$, $B_0$, $E_{\mathrm{NBI}}$, $\langle n \rangle$, $B_0$, and $P_{\mathrm{NBI}}$ were defined. From these parameters, most other major parameters in the plasma can be defined. Fixed parameters were $B_m = 25~\mathrm{T}$ and $\ell = 4.5~\mathrm{m}$. From the POPCON analysis in Section~\ref{sec:POPCONs}, the acceptable values of $\langle n \rangle$ were known to range from $1 \times 10^{20}~\mathrm{m}^{-3}$ to $2 \times 10^{20}~\mathrm{m}^{-3}$. By fixing $\langle n \rangle$ and $E_{\mathrm{NBI}} \sim E_i$ to some desired values, it is possible to presume that the electron energy is $\sim 0.1 E_{\mathrm{NBI}}$ which \citet{Egedal2022,Endrizzi2023} have shown is roughly accurate for an NBI heated simple mirror plasma plasma. The beam path was estimated to be $\sim 2\sqrt{2} a_m$ and the plug volume was taken to be $V \sim 2/3 \pi R_m a_m^2 \ell$ where the factor of 2/3 is added based on simulation results to account for the field curvature reducing the volume from a simple cylinder. With these assumptions in place, it was possible to constrain the rest of the mirror's parameters by simultaneously solving system:
\begin{subequations}\label{eq:constraints}
    \begin{align}
        &\underline{\text{NBI Absorption}} \notag \\
        &\exp{\left(-\int n \sigma_s dl\right)} < 0.2 \label{eq:NBI_abs}\\
        &\underline{\text{DCLC Stability}} \notag \\
        &\frac{a_0}{\rho_i} \gtrsim 25 \label{eq:DCLC}
    \end{align}
\end{subequations}
for $a_m$ and $B_0$. The NBI absorption condition \eqref{eq:NBI_abs} uses the analytic NBI stopping cross sections $\sigma_s$ from \citet{Janev1989} and \eqref{eq:DCLC} is a slightly modified version of the high-$\beta$ drift-cyclotron wave stability criterion which appears in \citet{Baldwin1977} and \citet{Tang1972}. According to these conditions, cool plasma trapped by sloshing ions \textit{or} high-$\beta$ and large plasma size can be used to stabilize a plasma against DCLC. In the end plug here, stabilization of DCLC will come not from sloshing ions, but primarily from large plasma size and high-$\beta$ as the POPCONs in Section~\ref{sec:POPCONs} predict that $T_e$ will be too high for effective sloshing ion trapping of cool plasma due to $\tau_{ii} \ll \tau_{d}$ (This stabilization method has been demonstrated experimentally in the LAMEX mirror \citep{ferron1984}). Using the assumptions outlined above, \eqref{eq:constraints} may be rewritten (taking \eqref{eq:DCLC} to be equal to 25):
\begin{subequations} \label{eq:constraints_v2}
    \begin{align}
        &\underline{\text{DCLC Stability}} \notag \\
        &B_0^2 - 0.16\left(\frac{\mu E_{i,100\mathrm{keV}}}{B_m a_m^2} \right) - 3.0 E_{i,100 \mathrm{keV}} \langle n_{20} \rangle = 0 \\
        &\underline{\text{NBI Absorption}} \notag \\
        &6.7\times 10^{-21} =  \langle n_{20} \rangle \frac{\sqrt{1+\cot^2(\theta_{\mathrm{NBI}}/4)}}{\sin \theta_{\mathrm{NBI}}} a_m \sqrt{\frac{B_m}{B_0 (1-\beta)^{1/2}}} \sigma_s,
    \end{align}
\end{subequations}
where ion energies are in 100's of keV and densities are in $10^{20}~\mathrm{m}^{-3}$ (other units are standard MKS). At high values of $\langle n \rangle$ and $E_{\mathrm{NBI}}$, the constraint on $B_0$ set by $\beta$ rather than DCLC stability. In this case $B_0$ was constrained with condition $\beta \leq 0.5$. Once the constraint equations were solved, from the simple mirror confinement time scaling law $n_{20}\tau_p = 0.25 E_{\mathrm{NBI},100 \mathrm{keV}}^{3/2} \log_{10}(R_m)$ from \citet{Egedal2022} (this scaling law is for deuterium and does not include all the details of the scaling laws presented in \eqref{eq:logan_scaling}, but is sufficient for the constraint exercise here), one can calculate the upper bound on the absorbed $P_{\mathrm{NBI}}$ required to reach the target value of $\langle n \rangle$ using fueling rate balance:
\begin{equation}
    P_{\mathrm{NBI}} = \frac{E_{\mathrm{NBI}} \langle n \rangle V}{\tau_p}   
\end{equation}
From the POPCON analysis performed in Section~\ref{sec:POPCONs}, it was identified that plasma radius $a_m \gtrsim 0.15~\mathrm{m}$ and density $\langle n \rangle \sim $1.0-1.5$\times 10^{20}~\mathrm{m}^{-3}$. Plotting contours of $P_{\mathrm{NBI}}$ and $a_m$ versus $E_{\mathrm{NBI}}$ and $\langle n \rangle$ produces a plot like that shown in Figure~\ref{fig:constraints}. This plot indicates an operating region accessible with reasonable $P_{\mathrm{NBI}}$ values $<20~\mathrm{MW}$ exists between $E_{\mathrm{NBI}}$ of $200~\mathrm{keV}$ and $300~\mathrm{keV}$. However, the method used to constrain NBI absorption with \eqref{eq:NBI_abs} is suboptimal. It has not accounted for that, despite not being absorbed as readily at high energies, the beam may still be more efficient at fuelling, and simplified scaling laws were used here rather than an actual Fokker-Planck calculation. Furthermore, the electron temperature in this analysis is implicitly taken to be $\sim 0.1 E_\mathrm{NBI}$ in accordance with the scaling law from \citet{Egedal2022}. This is lower than the values predicted by the POPCONs in Section~\ref{sec:POPCONs}. These discrepancies will be corrected in the next parts of this section with more detailed simulations using the RealTwin. 

\subsection{Parametric simulations of standalone simple mirror performance}\label{sec:simp_mirror}

\begin{table} 
  \begin{center}
\def~{\hphantom{0}}
  \begin{tabular}{lll}
      Parameter  & Section~\ref{sec:simp_mirror} & Section~\ref{sec:selfconsist}\\
       $a_m$     & 0.15 - 0.25 m & 0.15 - 0.25 m \\
       $\ell_p$  & 4.5 m & 4.5 m\\
       $B_m$     & 25 T & 25 T\\
       $B_0$     & 4.0 - 6.0 T & 5.5 - 9.0 T \\
       $E_{\mathrm{NBI}}$ & 240 - 480 keV & 250 - 500 keV\\
       $P_{\mathrm{NBI}}$ & 10 - 20 MW & 10 - 20 MW\\
  \end{tabular}
  \caption{Mirror parameter ranges used in the parametric scan in Section~\ref{sec:simp_mirror} and in the machine learning aided optimization in Section~\ref{sec:selfconsist}.}
  \label{tbl:plug_sims}
  \end{center}
\end{table}

With the constraints in the previous section established, a parametric scan in $B_0$, $P_{\mathrm{NBI}}$, $E_{\mathrm{NBI}}$ and $a_m$ was carried out with the RealTwin model described in Section~\ref{sec:computational_model} was specified. This scan used the end plug parameters shown in Table~\ref{tbl:plug_sims} to simulate a standalone simple mirror. For this work, points in the scan were selected over a linear range for each of the parameters. This scoping method is coarse, but substantial constraints were already placed on the problem through the use of the POPCON analysis in Section~\ref{sec:POPCONs} and the analytic arguments used to constrain the physics and engineering earlier in this section. This made the coarse high fidelity scan sufficient to deliver an optimized simple mirror that exceeded the end plug requirements identified by the POPCON analysis. A scan in $\ell$ was also performed to determine if scaling power density with $\ell$ versus $P_{\mathrm{NBI}}$ was substantially different (changing the aspect ratio of the plug can slightly change NBI path length and modify NBI absorption). Varying $\ell$, however, had effectively the same impact on power density and plug performance as varying $P_{\mathrm{NBI}}$. In the rest of the discussion here will focus on the highest performing $\ell = 4.5~\mathrm{m}$ case.

The CQL3D-m simulations performed in this section used 24 flux surfaces evenly spaced from 0.01 to 0.9 in normalized square-root poloidal flux $\sqrt{\psi_n}$, and the axial $z$ grid used to formulate the ambipolar response along the field line had 64 evenly spaced points extending from the midplane to the mirror throat. The distribution functions were discretized over 500 points in momentum-per-rest-mass $u_0$ with a grid which used increased resolution at low energies as to accurate capture the ion distributions and 400 evenly spaced points in the pitch angle $\vartheta$. The maximum energy in the momentum-per-rest-mass grid was set to correspond to a $250~\mathrm{keV}$ electron energy. The CQL3D-m simulations used 800 timesteps. The time advance was initially set to $1~\mathrm{ms}$ for the first 200 time steps, then increased to $2~\mathrm{ms}$ for the next 200 time steps, and then further increased to $4~\mathrm{ms}$ for the remaining 400 timesteps providing an overall simulation runtime of $2.2~\mathrm{s}$, several $\tau_p$. Ramped time-stepping was used to maintain numerical stability, as despite being an implicit time-advance code, the iterations of CQL3D-m with Freya and Pleiades and the internal ambipolar potential solve are explicit. The Freya NBI deposition calculation was run at each CQL3D-m time step using $2.5\times10^6$ particles, and the Pleiades equilibrium was updated after every 100 time steps.

\begin{figure*}
  \centering
  \includegraphics[width=0.9\linewidth]{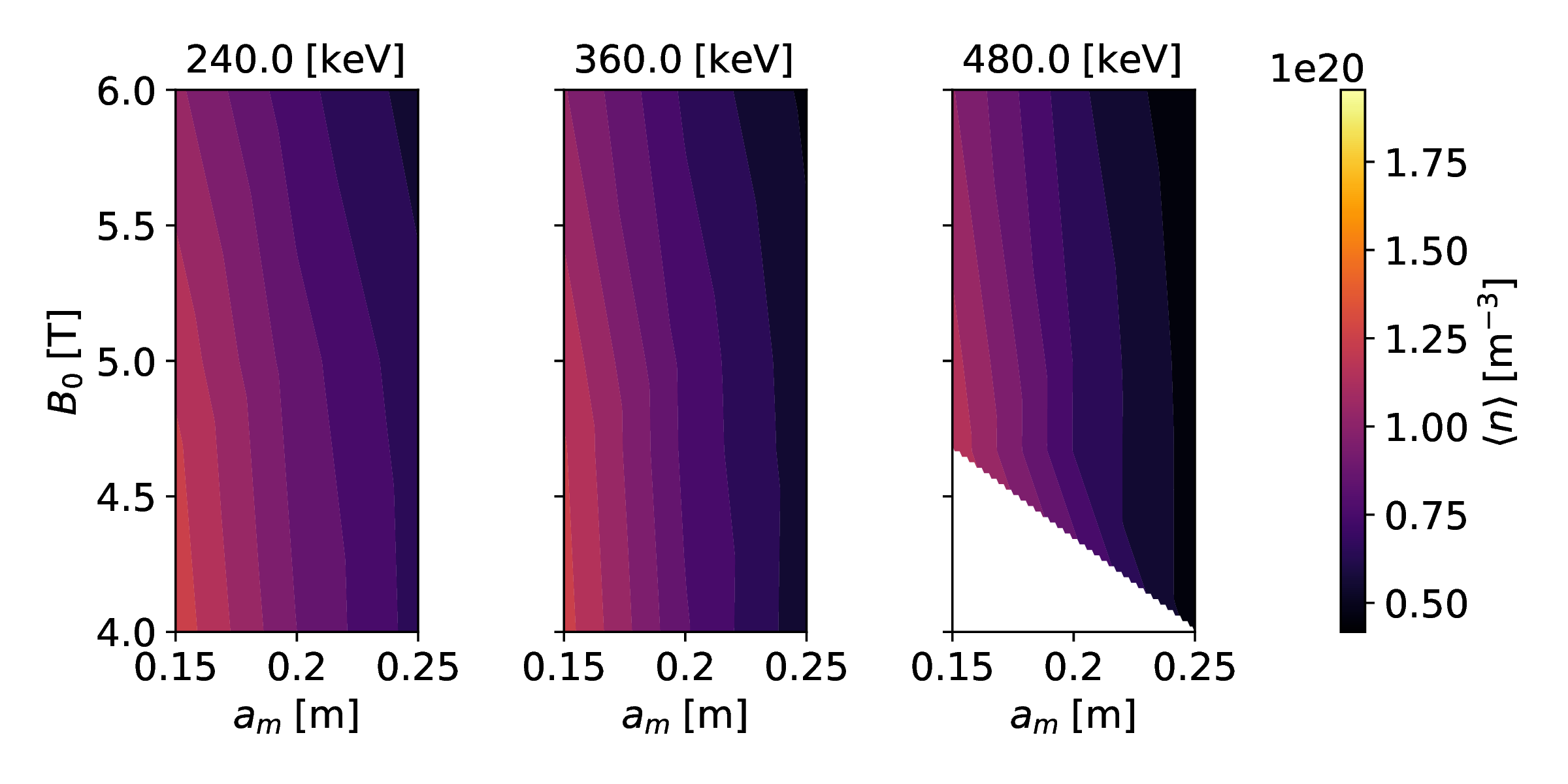}
  \includegraphics[width=0.9\linewidth]{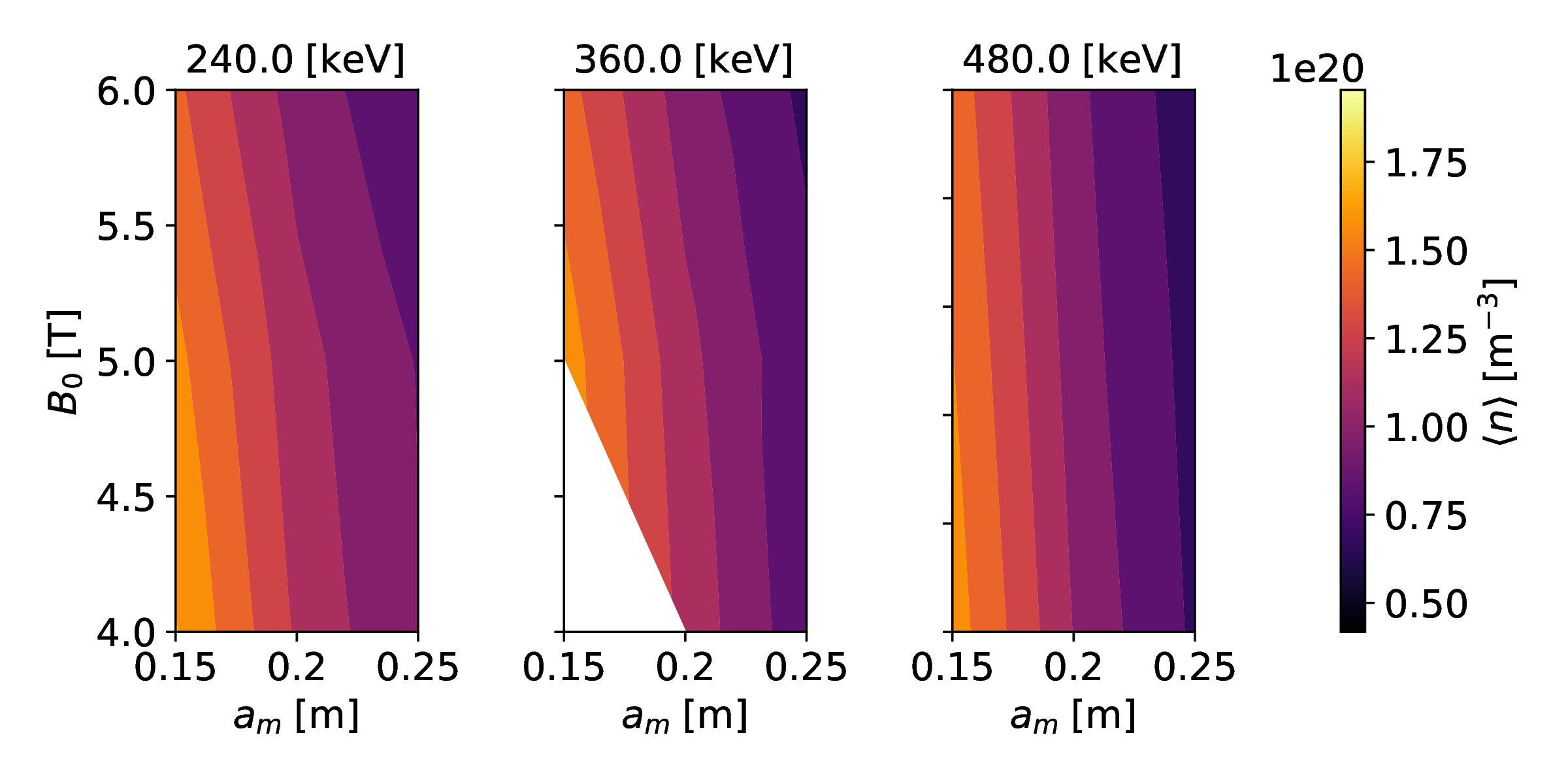}
  \includegraphics[width=0.9\linewidth]{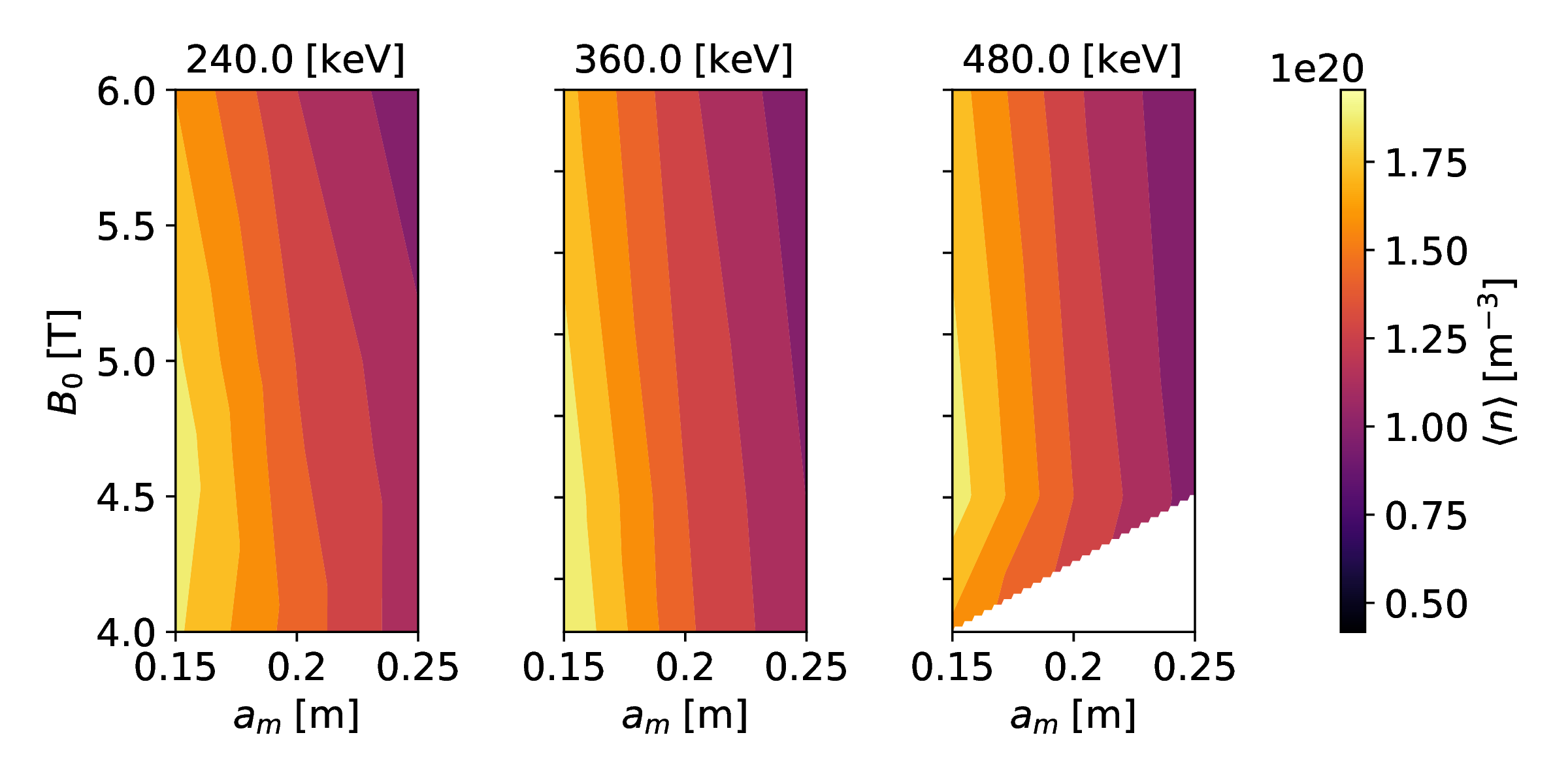}
  \caption{Contour plots of $\langle n \rangle$ for different values of $a_m$ and $B_0$ at three different $E_{\mathrm{NBI}}$ for an end plug with fixed length $\ell = 4.5~\mathrm{m}$ with three different NBI heating powers $10~\mathrm{MW}$ (top), $15~\mathrm{MW}$ (middle), and $20~\mathrm{MW}$ (bottom).}
\label{fig:rt_results}
\end{figure*}

The results of the simulations using the RealTwin model are shown in Figure~\ref{fig:rt_results}. Simulations which failed to satisfy the DCLC constraint \eqref{eq:DCLC} or in which the equilibrium failed to converge due to $\beta>1$ or the onset of the mirror instability were removed from this analysis (this is why there is an empty regions in the Figure~\ref{fig:rt_results} plots. The onset of the mirror instability as the source of a $\beta$ limit in mirror plasmsas will be discussed further in a future publication). A potentially surprising result of these scans is that there is little difference in peak density versus $E_{\mathrm{NBI}}$, however, remembering the analysis in Section~\ref{sec:assumptions}, this makes sense. Neutral beam absorption drops with higher $E_{\mathrm{NBI}}$, but the NBI fueling efficiency increases $\propto E_{\mathrm{NBI}}^{1/2}$. Another observation is that $\langle n \rangle$ improves superlinearly with increasing $P_{\mathrm{NBI}}$. This is due to a twofold non-linearity in the NBI absorption equation \eqref{eq:NBI_abs} with NBI power. As $P_{\mathrm{NBI}}$ increases, the $\langle n \rangle$ increases and therefore the neutral beam absorption fraction increases. In addition, the plasma $\beta$ also increases as $P_{\mathrm{NBI}}$ increases. This causes the plasma to expand outwards from diagmagnetism, increasing the NBI absorption path length, further increasing NBI absorption. 

\begin{figure*}
  \centering
  \includegraphics[width=0.9\linewidth]{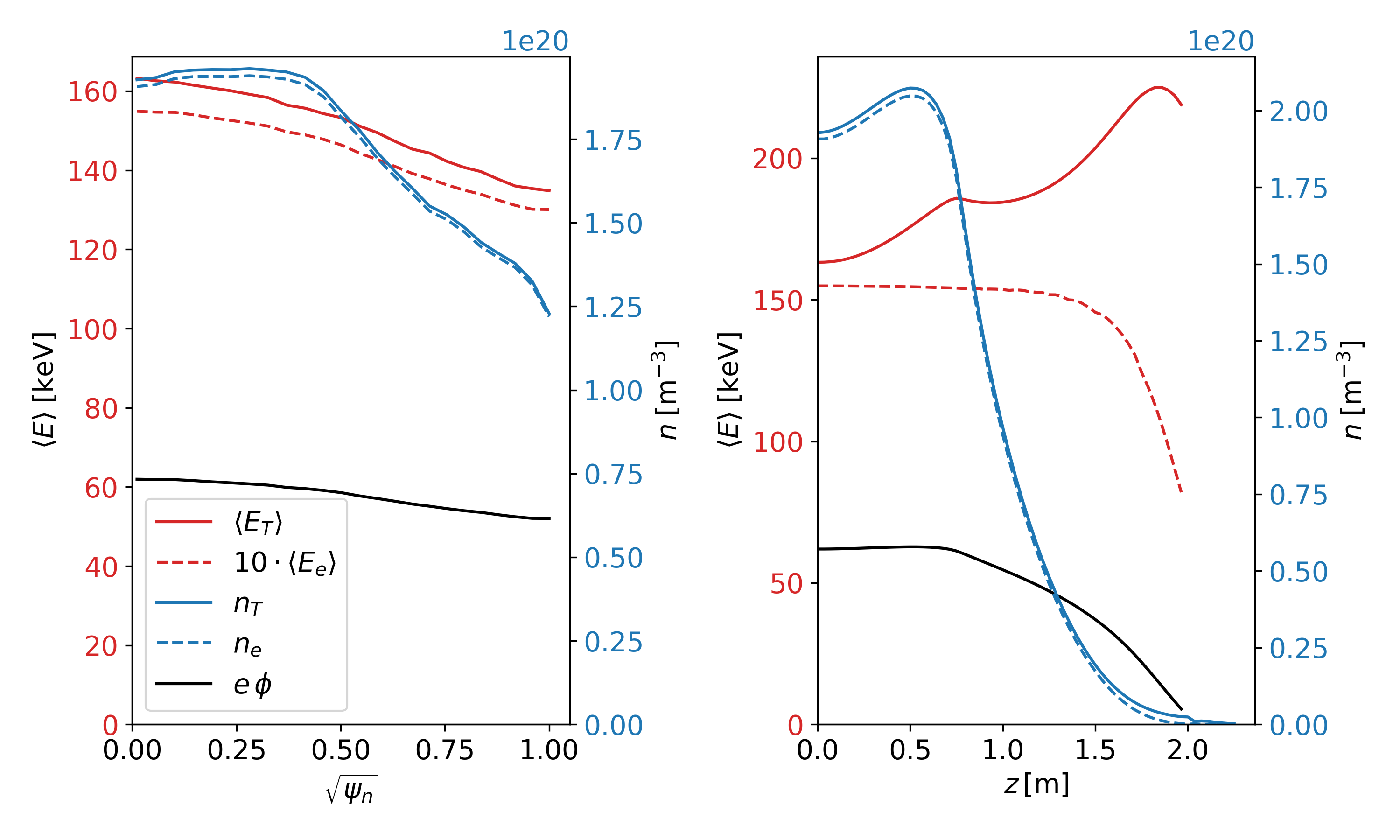}
  \caption{Density $n_p$ and average ion energy $\langle E_i \rangle$, electron energy $\langle E_e \rangle \equiv 1.5 T_e$, and ambipolar potential $e\phi$, versus the square root of the normalized poloidal flux $\sqrt{\psi_n}$. The profiles come from a simulation with parameters: $a_m = 0.15~\mathrm{m}$, $B_m = 25~\mathrm{T}$, $B_0 = 4~\mathrm{T}$, $\ell = 4.5~\mathrm{m}$, $E_{\mathrm{NBI}} = 240~\mathrm{keV}$, and $P_{\mathrm{NBI}} = 15~\mathrm{MW}$.}
\label{fig:plug_dens}
\end{figure*}

As anticipated in Section~\ref{sec:assumptions}, the target end plug $\langle n\rangle$ values identified in Section~\ref{sec:POPCONs} were achieved at the lowest $P_{\mathrm{NBI}}$ for $E_{\mathrm{NBI}}$ of $240~\mathrm{keV}$ and $360~\mathrm{keV}$. Sufficient end plug conditions for fusion relevant central cell operation were achievable with $B_0$ of $4 - 6~\mathrm{T}$, $P_{\mathrm{NBI}} \sim 15~\mathrm{MW}$, $E_{\mathrm{NBI}} \sim 240-360~\mathrm{keV}$, and $a_m$ of $0.15 - 0.2~\mathrm{m}$. Under these conditions effectively full NBI absorption (>95\%) was obtained with a $240~\mathrm{keV}$ beam, so lowering NBI energy further would only serve to reduce performance. The plasma profiles of the most optimal mirror identified in this scan are shown in Figure~\ref{fig:plug_dens}. The mirror had parameters: $a_m = 0.15~\mathrm{m}$, $B_0 = 4~\mathrm{T}$, $E_{\mathrm{NBI}} = 240~\mathrm{keV}$, and $P_{\mathrm{NBI}} = 15~\mathrm{MW}$ and achieved $\langle n \rangle > 1.5\times 10^{20}~\mathrm{m}^{-3}$, more than sufficient to drive a $50~\mathrm{m}$ long central cell to $Q > 5$ conditions based on the analysis in Section~\ref{sec:POPCONs}. 

\begin{table}
\begin{center}
\def~{\hphantom{0}}
  \begin{tabular}{lll}
      Parameter  & Optimum Standalone & Optimum Consistent $T_e$ \\
      \\
       \underline{End plug} \\
       Ion Species & T & D \\
       $a_m$     & 0.15 m & 0.25 m \\
       $a_0$     & 0.47 m & 0.59 m \\
       $\ell_p$  & 4.5 m & 4.5 m \\
       $B_m$     & 25 T & 25 T\\
       $B_{0p}$     & 4.0 T & 6.0 T\\
       $\langle n_{p0} \rangle$ & $1.66\times10^{20}$ m$^{-3}$ & $1.60\times10^{20}$ m$^{-3}$ \\
       $\langle\beta_{p0}\rangle$ & 0.58 & 0.59\\
       $E_{\mathrm{NBI}}$ & 240 keV & 403 keV \\
       $T_e$ & 11 keV & 120 keV \\
       $P_{\mathrm{NBI}}$ & 15 MW & 20 MW \\
       $\theta_\mathrm{NBI}$ & $45^\circ$ & $50^\circ$\\
       $\langle a_0 / \rho_{0i} \rangle$ &  24 & 29
       \\
       \underline{Central cell} \\
       Ion Species & DT (50:50) & DT (50:50) \\
       $a_c$     & 0.54 m & 0.86 m \\ 
       $l_c$     & 50 m & 50 m \\
       $B_{0c}$  & 3.125 T & 3.125 T\\
       $\beta_c$ & 0.6 & 0.55 \\
       $n_c$     & $7.5\times10^{19}$ m$^{-3}$ & $7.0\times10^{19}$ m$^{-3}$ \\
       $T_{ic}$  & 50 keV & 57 keV \\
       $T_{ec}$  & 100 keV & 120 keV \\
       $P_{\mathrm{fus}}$ & 175 MW & 350 MW \\
       $Q$       & 5.8 & 8.75 \\
       
  \end{tabular}
  \caption{Optimized tandem mirror parameters based on the results of parametric scans and POPCON analysis using the end plug conditions obtained using the standalone simulations with CQL3D-m and Pleiades performed in Section~\ref{sec:simp_mirror} in column "Optimum Standalone" and the CQL3D-m and Pleiades simulations with electron temperature fixed to the central cell value performed in Section~\ref{sec:selfconsist} in column "Optimum Consistent $T_e$".}
  \label{tbl:opt_params}
  \end{center}
\end{table}

\subsection{Optimization for a self-consistent end plug operating condition}\label{sec:selfconsist}

While the end plug-like simple mirror identified in Section~\ref{sec:simp_mirror} appears to be sufficient to deliver a viable central cell, the scan those conditions were derived from was coarse. It also did not include the effects of increased electron temperature in the end plug as a result of alpha particle heating from the central cell. To address this, a secondary set of RealTwin simulations were created. These simulations used a modified CQL3D-m model in an attempt to better capture the tandem mirror physics. The electron distribution was assumed to be fixed by the classical tandem mirror central cell dynamics and the CQL3D-m simulations in this section used a fixed $T_e$ Maxwellian electron background where $n_e=n_i$ and a fixed ambipolar potential $\phi_e$ corresponding to those computed in the tandem mirror POPCON using \eqref{eq:popcon_sys} and \eqref{eq:phie}. The operating point for the tandem mirror was chosen to be at a thermally stable ignited central cell (i.e. operating on the right hand side of the ignited region) operating point with $\beta_c \sim 0.6$ with the same plug density as the RealTwin simulations. The POPCON and the RealTwin simulations were iterated until a self-consistent condition where $T_e$, $\phi_e$, and $\langle n_p \rangle$ in the POPCON and RealTwin models matched. These RealTwin simulations consistent with the POPCON conditions provide a more detailed picture of steady-state end plug operation in a tandem mirror. Most notably, the larger ambipolar hole in the ion distribution increases the optimal values of $E_\mathrm{NBI}$ and $B_0$.

In addition to fixed $T_e$ and $\phi_e$, these simulations employed several other modifications and refinements on the model versus Section~\ref{sec:simp_mirror}. Rather than the coarse parameteric scan used in the previous section, the Bayesian optimization technique described in Section~\ref{sec:computational_model} was used to efficiently find values of  $a_m,$ $B_0,$ $E_\mathrm{NBI},$ and $P_\mathrm{NBI}$ which optimize the normalized Lawson criterion $\langle n \tau \rangle (P_\mathrm{abs}/P_\mathrm{NBI})$, where $P_\mathrm{abs}/P_\mathrm{NBI}$ is the absorbed fraction of the total NBI power. This metric can be thought of as a measure of NBI fuelling efficiency. The normalized Lawson criteria is the preferred objective function quantity as it simultaneously optimizes density per unit of absorbed NBI power and ensures good beam absorption. Rather than the simple $a_0/\rho_{i0} > 25$ DCLC constraint used previously, the $\beta$ adjusted DCLC constraint from \citet{Tang1972}
\begin{equation}\label{eq:tang}
    N_\rho^{\mathrm{Tang}} \equiv a_0/\rho_{i0} \approx \frac{1}{2}\frac{1}{\sqrt{\beta}\sqrt{m_e/m_i + \Omega^2_i /\omega_{pi}^2}},
\end{equation}
was employed in order to provide a more accurate estimate of the DCLC stability. This constraint was implemented automatically in the objective function through the use of penalty function: $f(a_0/\rho_i) = 1/2 \tanh{\{[a_0/\rho_i - (N_\rho^{\mathrm{Tang}} - 5)]/5\}}$ which multiplies $\langle n \tau \rangle (P_\mathrm{abs}/P_\mathrm{NBI})$. 

The optimization in this section was initially performed using tritium to improve classical confinement. However, the higher $E_\mathrm{NBI}$ required to overcome the ambipolar hole at high $T_e$ caused end plugs using tritium as the main ion species to not satisfy the DCLC constraint. Because of this, a switch to deuterium was made to reduce $\rho_i$. The confinement reduction factor resultant from this switch was $\sim\sqrt{3/2}$ rather than $\sim3/2$ as the elevated $T_e$ in these simulations caused the plasma to be ion-ion scattering dominated rather than electron drag dominated as it was in Section~\ref{sec:simp_mirror}. 

In order to recover some classical confinement performance in deuterium operation, the NBI angle $\theta_\mathrm{NBI}$ was shifted from the value of $45^\circ$ used in Section~\ref{sec:simp_mirror} to $50^\circ$, increasing the effective mirror ratio. This small shift was acceptable as $\theta_\mathrm{NBI}$ should not be perpendicular enough to excite the Alfven ion-cyclotron instability \citep{Smith1984} or the mirror instability \eqref{eq:mirrorinstab}, sloshing ions are not required for DCLC stability, and the NBI path length reduction is small enough that the shift in $\theta_\mathrm{NBI}$ does not significantly reduce ${P_{abs}}/{P_\mathrm{NBI}}$.

\begin{figure*}
  \centering
  \includegraphics[width=0.6\linewidth]{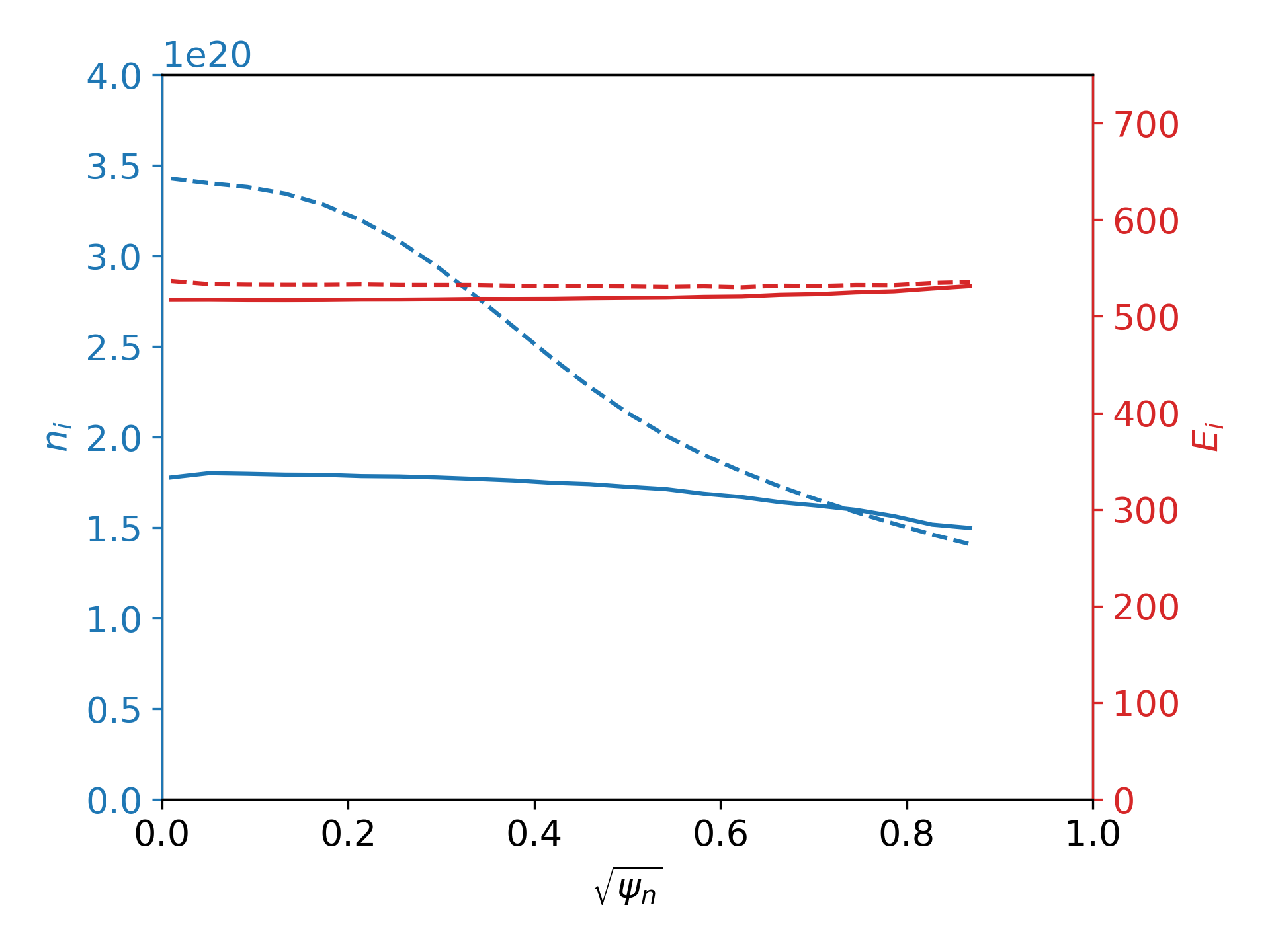}
  \caption{Plots of density $n$ (blue) and ion temperature $E_i$ (red) versus normalized square root poloidal flux $\sqrt{\psi_n}$ obtained from CQL3D-m simulations of a fixed $\beta=0.6$ end plug with $a_m=0.25~\mathrm{m}$, $B_m = 25~\mathrm{T}$, $B_0 = 6~\mathrm{T}$, $E_\mathrm{NBI} = 400~\mathrm{keV}$, $P_\mathrm{NBI} = 20~\mathrm{MW}$, and $T_e = 120~\mathrm{keV}$ both with (solid) and without (dashed) classical radial diffusion. Radial diffusion has a significant affect on confinement in these cases. Diffusion notably reduces the total density in addition to flattening the density. The $E_i>E_\mathrm{NBI}$ seen in the plot is the result of a combination of up-scattering in ion energy and the preferential loss of low energy ions.}
\label{fig:rad_diff}
\end{figure*}

End plug simulations, especially those with large $E_\mathrm{NBI}$, were found to exhibit an instability tied to $\beta \rightarrow 1$ near the magnetic axis. The higher NBI power density due to smaller flux surface volumes as $\sqrt{\psi_n}\rightarrow0$ (low $\sqrt{\psi_n}$ surfaces absorb a similar amount of NBI power to large $\sqrt{\psi_n}$ surfaces for an NBI width substantially less than $a_0$) caused $\beta$ on the inner flux surfaces to grow quickly. Higher $\beta$ increased $R_m$ and $\tau_p$. This in turn increased $n$ leading to a further increase to the NBI power density on the inner flux surfaces. This feedback loop rapidly caused $\beta$ on the inner flux surfaces to exceed unity resulting in a loss of magnetic equilibrium. To combat the instability, a radially varying classical diffusion coefficient based on \eqref{eq:tauclassical} was added to CQL3D-m's Fokker-Planck equation in order to flatten the pressure profiles in the end plug. A typical value for this diffusion coefficient is $D_\rho \sim 1\times10^{-4}~\mathrm{m^2s^{-1}}$. This addition caused a reduction in performance due to the introduction of radial losses, but classical diffusion successfully resolved the $\beta \rightarrow 1$ instability in most cases and flattened the radial end plug density profiles. An example of the profile flattening from radial diffusion is shown in Figure~\ref{fig:rad_diff}. 

However, when the magnetic equilibrium was evolved self-consistently with Pleiades using the pressure profiles from the CQL3D-m simulations with diffusion, numerical noise developed which could corrupt the solution. The noise appeared to be related to the radial boundary conditions used in CQL3D-m, and there will be attempts to resolve this noise in future work so that fully self-consistent CQL3D-m/Pleiades optimizations of end plug performance may be performed. To combat the noise here, fixed $\beta_p = 0.6$ Pleiades equilibria were used during the optimization to prevent failures which can impair the GPBO's convergence. To maintain a degree of self-consistency, the optimization imposed a Gaussian envelope penalty function (centred at $\beta_p = 0.6$ with $\sigma = 0.1$) multiplying $\langle n \tau \rangle (P_\mathrm{abs}/P_\mathrm{NBI})$ to ensure that optimized end plugs had $\beta_p\sim0.6$. After optimal end plug conditions were identified, the optimal cases were rerun in standalone simulations which self-consistently included field evolution to confirm that the same results could be obtained.

The CQL3D-m simulations in this section could be evolved for a longer period of time at lower resolution as the explicit potential solver was disabled and electrons were considered as a background species rather than undergoing nonlinear evolution with the ions. These CQL3D-m simulations used 24 flux surfaces in the radial grid even spaced from $\sqrt{\psi_n} = 0.01-0.95$ and 64 axial points evenly spaced in $\hat z$ between the midplane and the mirror throat. The distribution function was discretized over 300 points in momentum-per-rest-mass $u_0$ and 200 evenly spaced points in the pitch angle $\vartheta$. The maximum energy in the momentum-per-rest-mass grid was set to correspond to $1000~\mathrm{keV}$ deuterons. CQL3D-m used 100 ms timesteps and Freya NBI deposition calculations were run at each time step using $2.5\times10^6$ particles. Simulations were evolved for $10~\mathrm{s}$, several $\tau_p$, to time converge the end plug transport. Rather than self-consistently optimize both the POPCON and the end plug in a single simulation loop, optimal end plug conditions were calculated at a number of different values of $T_e$ independently using the Bayesian optimization procedure described in Section~\ref{sec:computational_model}. At each value of $T_e$, five batches of 10 simulations were run with parameters selected using the CL-min acquisition function. The initial three batches used a more exploratory acquisition function with $\kappa = 10$ and final two batches used a lower $\kappa = 5$ which allowed the optimization to quickly find a global optimum.

\begin{figure*}
  \centering
  \includegraphics[width=0.6\linewidth]{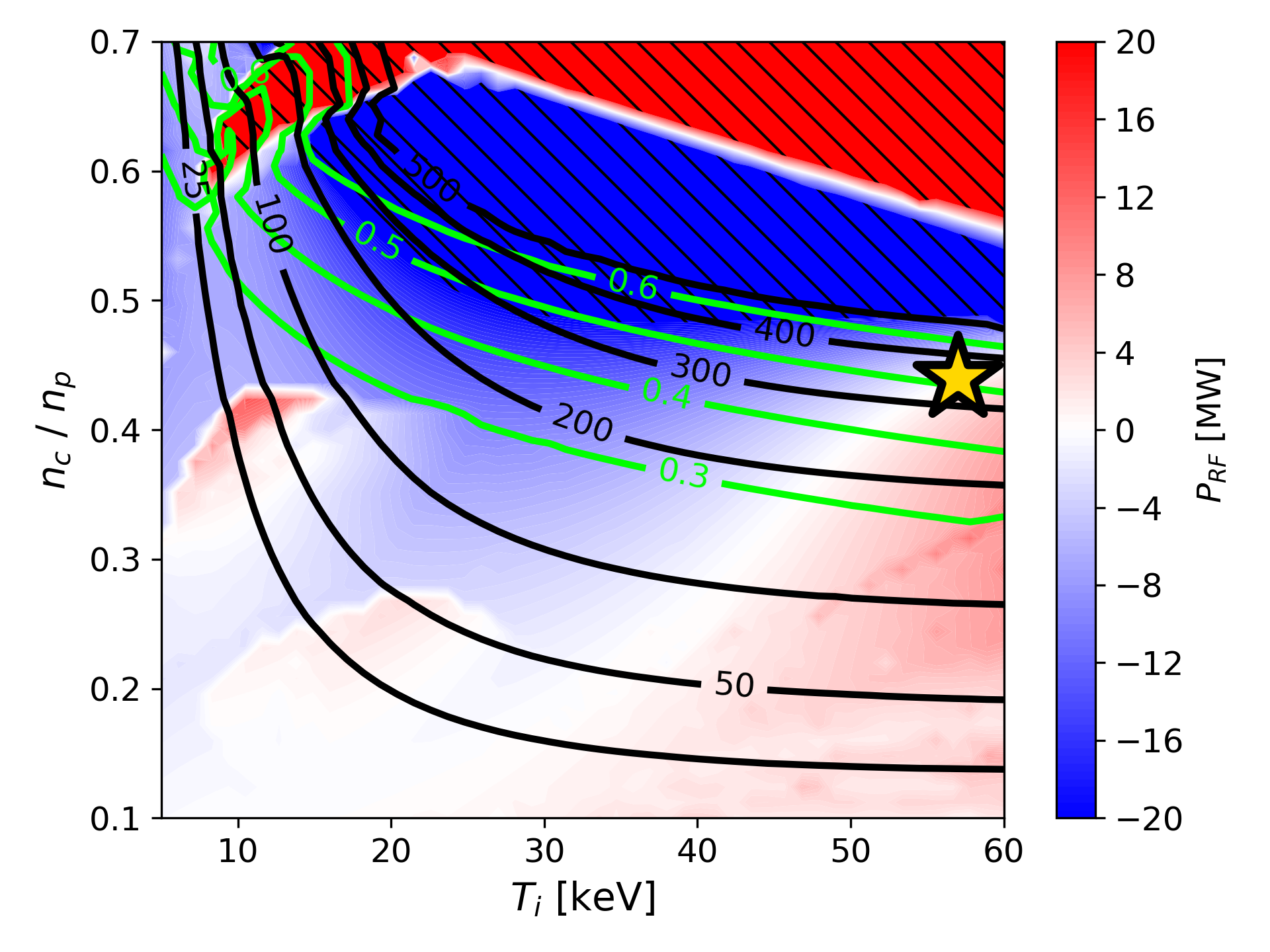}
  \caption{A POPCON of the tandem mirror operating point described the second column labelled "Optimum Consistent $T_e$" Table~\ref{tbl:opt_params}. Heating power $P_{\mathrm{RF}}$ is shown in the filled red (positive) and blue (negative, ignited) contours, fusion power $P_{fus}$ in black contours, $\beta_c$ in green contours, and the operating point in the table is marked with a star.}
\label{fig:opt_endplug}
\end{figure*}

As anticipated in Section~\ref{sec:simp_mirror}, when self-consistent electron temperatures were accounted for, $E_\mathrm{NBI}$, $B_0$, and $a_m$ had to increase to ensure that the plasma remained below the $\beta$ limit, had good confinement, had adequate NBI absorption, and remained stable to DCLC. The optimization procedure selected the largest end plug radii $a_m=25~\textrm{cm}$ set by engineering feasibility as it both improved $\langle n\tau_p\rangle (P_\mathrm{abs}/P_\mathrm{NBI})$ and allowed the plasma to more easily satisfy the DCLC stability condition \eqref{eq:tang}. With end plug $T_e = 80-120~\mathrm{keV}$, $E_\mathrm{NBI} \approx 400~\mathrm{keV}$, and $P_\mathrm{NBI} \approx 15-20~\mathrm{MW}$, desirable $1.5\times10^{20}~\mathrm{m}^{-3}$ densities were achievable. These densities were achievable despite the $a_m = 25~\mathrm{cm}$ plugs' higher volume because of the increased fuelling efficiency obtained with high-energy NBI and the reduction in electron drag from using self-consistent values for $T_e$. The best end plug performance was obtained at $T_e=100~\mathrm{keV}$ as it provided a good balance of increased ambipolar losses and decreased electron drag. However, for the best overall system performance occurred at $T_e=120~\mathrm{keV}$, and it was chosen for the optimized operating point as it increased the achievable central cell $P_{fus}$ by $\sim 100~\mathrm{MW}$. The optimized tandem mirror POPCON operating point based on this optimization process is shown in Figure~\ref{fig:opt_endplug} and the parameters of the optimized tandem mirror are given in the second column labelled "Optimum Consistent $T_e$" Table~\ref{tbl:opt_params}. It is noted that, in future evolutions of this design, slightly higher $\beta$ may be desirable in order to achieve high $\beta$ stabilization using conducting walls \citep{Berk1987,Li1987} and higher $P_\mathrm{fus}$. This may be achieved under the same parameters with elevated $P_\mathrm{NBI}$, or can be optimized for in a new optimization. The optimized tandem mirror operating point in Table~\ref{tbl:opt_params} has self-consistent $T_e$, uses technologically feasible components, obeys the DCLC stability constraint \eqref{eq:tang} (the Tang constraint specified $a_0/\rho_i > 27$ for diamagnetic drift wave stability), and satisfies the NASEM requirements for a fusion power core \citep{Hawryluk2021}.

\section{Conclusions and future work}\label{sec:conclusion}

A new computational model for magnetic equilibrium, transport, heating, and fuelling in simple mirrors was developed, a novel POPCON technique for tandem mirror performance was derived, and optimized end plugs that deliver tandem mirror central cell performance consistent with the fusion pilot plant criteria outlined in the NASEM recommendations in the ``Bringing Fusion to the US Grid'' report were identified. The analysis here has identified an optimized $25~\mathrm{T}$ peak field tandem mirror end plug that achieves average densities in excess of $1.5 \times 10^{20}~\mathrm{m}^{-3}$ with $15~\mathrm{MW}$ of $240-360~\mathrm{keV}$ NBI heating and enables a $50~\mathrm{m}$ long $3~\mathrm{T}$ central cell to generate over $150~\mathrm{MW}$ fusion power. This operating point is significantly more conservative than previous classical tandem mirror concepts proposed in the literature \citep{Fowler1977,Fowler2017}. Along the way, a number of important new and important insights about classical axisymmetric tandem mirror operation were brought to light. Using the novel POPCONs approach derived here, it was found found that in a classical tandem mirror the end plug can be characterized sufficiently to estimate central cell performance with only parameters: the plasma radius at the mirror throat $a_m$,  the field at the mirror throat $B_m$, and the end plug density $n_p$. The POPCONs found that alpha particle heating in the central cell can substantially raise $T_e$ and is important to understanding tandem mirror performance, and the central cell density ratio $n_c/n_p$ in high-field axisymmetric tandem mirrors should be significantly higher than previously suggested in the literature. It was also noted that synchrotron radiation and classical transport could substantially limit tandem mirror central cell performance when $\beta$ becomes large, and therefore these effects must be included in the tandem mirror scopings. End plug performance and engineering feasibility were also found to be enhanced by fuelling the end plug with only tritium provided the DCLC stability conditions could be satisfied. Machine learning aided optimization of end plug equilibria and transport using integrated modelling was performed using Realta's RealTwin integrated model. It was shown that pilot plant relevant end plug parameters can be reached with near-term technologies. These simulations also revealed two important pieces of end plug physics that have been overlooked in previous studies such as \citet{Fowler2017}. First, NBI energy optimizes at a significantly lower value, $\sim 400~\textrm{keV}$, than what was used in previous tandem mirror scoping studies, $\sim 1~\textrm{MeV}$, when neutral beam shine through and DCLC are self-consistently accounted for. Second, classical transport is very important to moderating the value of $\beta_0$ at the magnetic axis of the end plug. Without diffusion, $\beta_0(\psi =0)$ in the end plug will rapidly approach unity for realistic NBI absorption profiles limiting the maximum value of $\langle n_p\rangle$. Furthermore, the flattening of end plug density profiles from diffusion could suppress drift waves, having a stabilizing effect on DCLC.

Along with these new discoveries, this work has exposed a number of important computational and theoretical capability gaps which need to be filled to further the modelling of tandem mirrors the most obvious of which is kinetic and MHD stability which were not considered rigorously here. Most stability actuators, such as shear flow, entail nonlinear stabilization processes that are difficult to resolve in stimulation and will require new nonlinear stability models like that shown in \citet{Tran2024} coupled with the equilibrium models presented here. In addition to the limitations associated with predicting stability the equilibrium models here the piece-wise approach to estimate central cell performance based on end plug performance required a number of assumptions. Ideally, a fully integrated model of the tandem mirror transport would be developed in the future. This could entail either a 2 real-space dimension 2 velocity-space dimension (2D2V) time-dependent Fokker-Planck solver or a set of bounce-averaged 1D2V time dependent Fokker-Planck solvers coupled together with the proper boundary-conditions and source/sink terms. A 2D2V FP solver would have the added benefit of resolving many of the issues which can arise with the ambipolar potential calculation on the addition of EC heating as a more accurate method for calculating the ambipolar potential such as that proposed in \citet{Degond2010} or \citet{Kupfer1996} could be used. Such a solver would also allow accurate consideration magnetic expander regions and the investigation of thermal barrier tandems, which could provide a large performance enhancement over the classical tandems investigated here. While it was found that the applicability of an RF ion acceleration scheme analogous to that in WHAM is limited in an end plug (see Appendix~\ref{sec:rf_limits}), further work on RF heating in magnetic mirrors is warranted. This would include the development of axisymmetric mirror geometry full-wave solvers and more innovative RF schemes which reduce quasilinear particle losses. Finally, while of limited importance in a negative-ion neutral beam fueled end plug, the impact of halo neutrals generated by NBI charge exchange and edge neutral charge exchange induced loss of fast ions are anticipated to be of some importance in smaller mirrors like WHAM. To better understand these effects the modelling tools used in this study will be integrated with edge neutral modelling codes.

\section*{Acknowledgements}
This research was funded by Realta Fusion and performed as part of the DOE Milestone Based Fusion Development Program DE-SC0024887. The authors would like to thank the DOE Milestone review committee for their helpful input which improved this paper. This research also used resources of the National Energy Research Scientific Computing Center, a DOE Office of Science User Facility supported by the Office of Science of the U.S. Department of Energy under Contract No. DE-AC02-05CH11231 using NERSC award FES-ERCAP0026655. Realta Fusion is a fusion energy company spun out of the ARPA-E funded WHAM project at the University of Wisconsin–Madison. Realta is developing compact magnetic mirror fusion power plants as a low capital cost path to fusion power, and targeting initial fusion power plant deployment at the $100-200~\mathrm{MW}$ scale at industrial sites and datacenters for process heat and/or electricity. 

\appendix

\section{Limitations of RF ion heating in end plugs}\label{sec:rf_limits}

In previous high-field mirror literature, there has been some discussion of RF heating end plug ions to improve particle confinement \citep{Forest2021,Endrizzi2023}. However, it was decided to not use RF heating in this study. Over the course of this work, it became clear that an RF ion heating scheme like that employed in WHAM \citep{Endrizzi2023}, in which positive-ion, $\sim100~\mathrm{keV}$, NBI ions are accelerated by $n=2-3$ harmonic ion-cyclotron heating, is not effective enough at increasing ion particle confinement, and at tandem mirror electron temperatures, positive-ion neutral beams with a beam energy of $120~\mathrm{keV}$ and an injection angle of $45^\circ$ will, according to \eqref{eq:ambi_loss_e}, be nearly immediately lost to the ambipolar potential. In addition, particle losses from enhanced quasilinear diffusion into the loss-cone eliminate much of the gain in particle confinement that can be obtained from increasing the ion energy. While ion energy confinement and fusion rates are increased with the addition of RF heating, this is of negligible importance to overall tandem mirror fusion performance (and enhanced fusion rates in the end plug are in fact undesirable in most circumstances). In the remainder of this appendix, a representative RF heating case in at end plug conditions will be analysed which demonstrate these challenges.

The simulation under consideration here utilizes a tritium-fueled end plug with $a_m = 0.15~\mathrm{m}$, $B_m = 25~\mathrm{T}$, $B_0 = 5~\mathrm{T}$, $\ell = 4.5~\mathrm{m}$, $E_{\mathrm{NBI}} = 120~\mathrm{keV}$, $P_{\mathrm{NBI}} = 10~\mathrm{MW}$, $f_{\mathrm{RF}} = 115.5~\mathrm{MHz}$, and $P_{\mathrm{RF}} = 5~\mathrm{MW}$. This end plug condition is similar to the optimized end plugs identified in Section~\ref{sec:optimization}, but rather than using a $15~\mathrm{MW}$ of negative-ion NBI power, it uses $15~\mathrm{MW}$ combined of positive-ion NBI and $n=3$ harmonic RF power. The RF antenna in these simulations was placed slightly beyond the fast ion turning point at $z = 1~\mathrm{m}$ calculated by CQL3D-m (this does not correspond exactly to $B/B_0 = 2$ due to the use of realistic neutral beam geometry and the effects of finite $\beta$) at the limiting flux surface $\sqrt{\psi_n} = 1.0$, and the RF source frequency $f_{\mathrm{RF}}$ was set to correspond to a location just inside the fast ion turning point. The antenna spectrum used in GENRAY-c was consistent with the loop style of antennas which will be tested on WHAM \citep{Endrizzi2023} and had a poloidal mode spectrum centred about $n_\theta = 0$ with a spectral width of $n_\theta \pm 1$ and a uniform axial mode spectrum $n_\phi = 0 - 10$ (this definition of the axial mode spectrum with $n_\phi$ is an artifact of the toroidal GENRAY code, but has been preserved in the text to make future replication with GENRAY-c easier). The simulations of this end plug were initially run with only NBI and no additional RF heating. CQL3D-m was iterated with Pleiades 8 times with 800 total CQL3D-m timesteps for $\sim 1~\mathrm{s}$ using only NBI heating until it reached steady-state. This was done in order to develop a target plasma with high enough $\beta$ to enable efficient cyclotron harmonic damping of the RF waves. When steady-state was achieved, GENRAY-c simulations were added to the CQL3D-m/Pleiades IPS workflow using self-consistent equilibrium and plasma profiles and coupled to CQL3D-m using an ion and electron quasilinear diffusion model. After the addition of RF, the simulations were run for an additional 1000 CQL3D-m time-steps, iterating Pleiades, CQL3D-m and GENRAY-c 8 times, for $\sim 1~\mathrm{s}$.

\begin{figure*}
  \centering
  \includegraphics[width=0.5\linewidth]{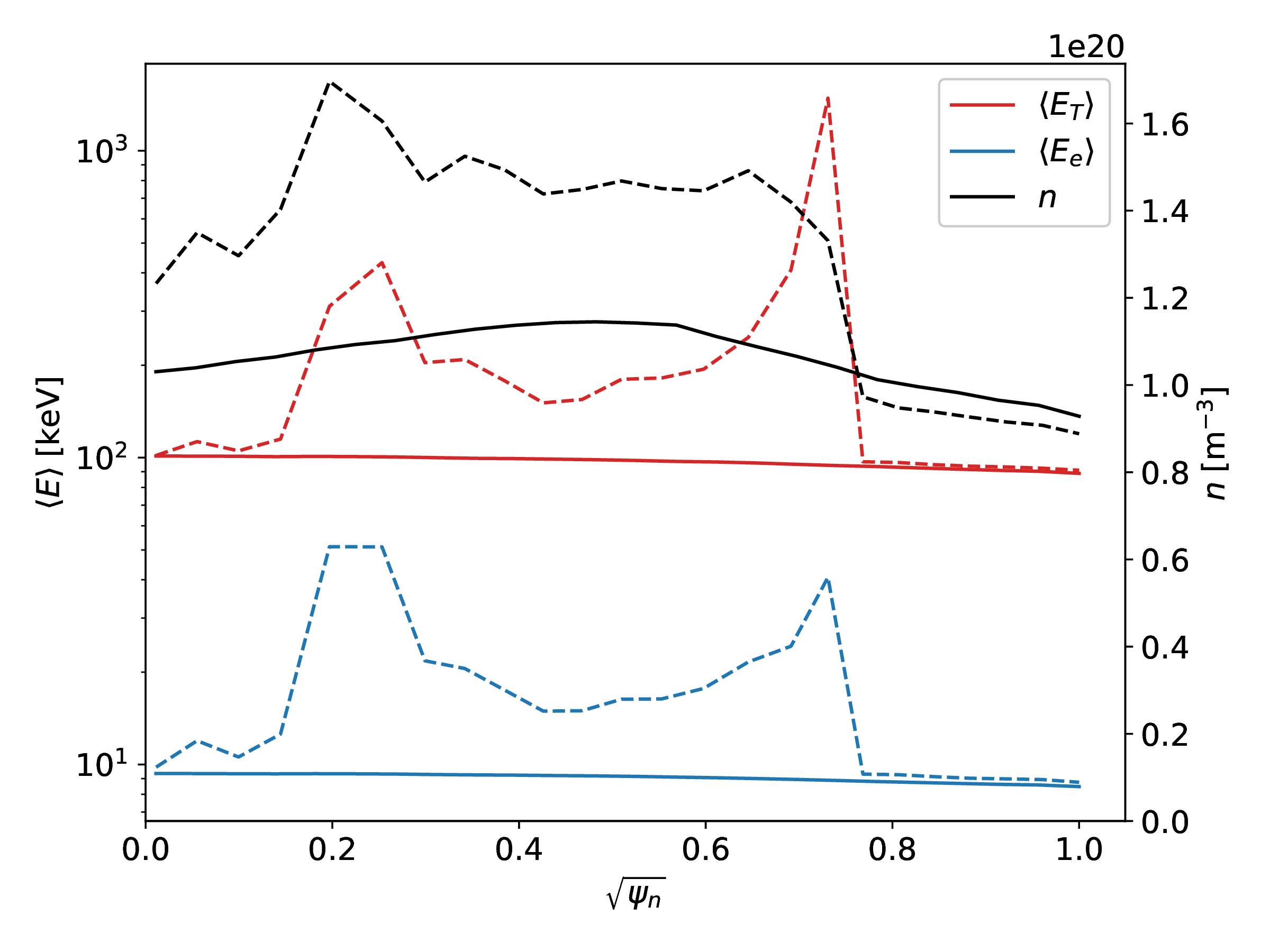}
  \includegraphics[width=0.475\linewidth]{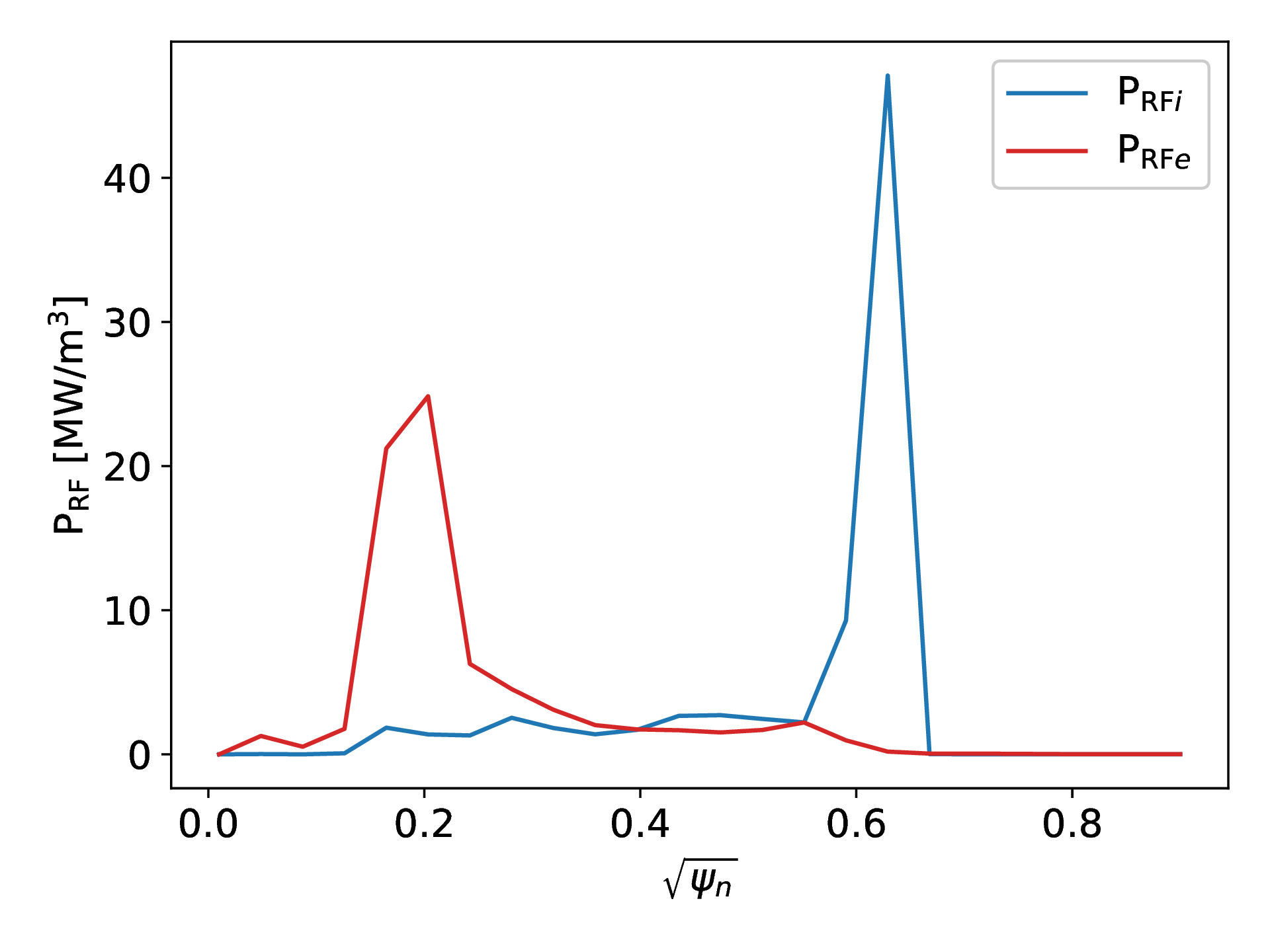}
  \caption{On the left ion density (blue) and temperature (red) profiles versus the normalized square root poloidal flux $\sqrt{\psi_n}$ for a simulation with RF (solid) and without RF (dashed), and on the right the deposited RF power density profiles versus $\sqrt{\psi_n}$.}
\label{fig:rf_density}
\end{figure*}

\begin{figure*}
  \centering
  \includegraphics[width=0.7\linewidth]{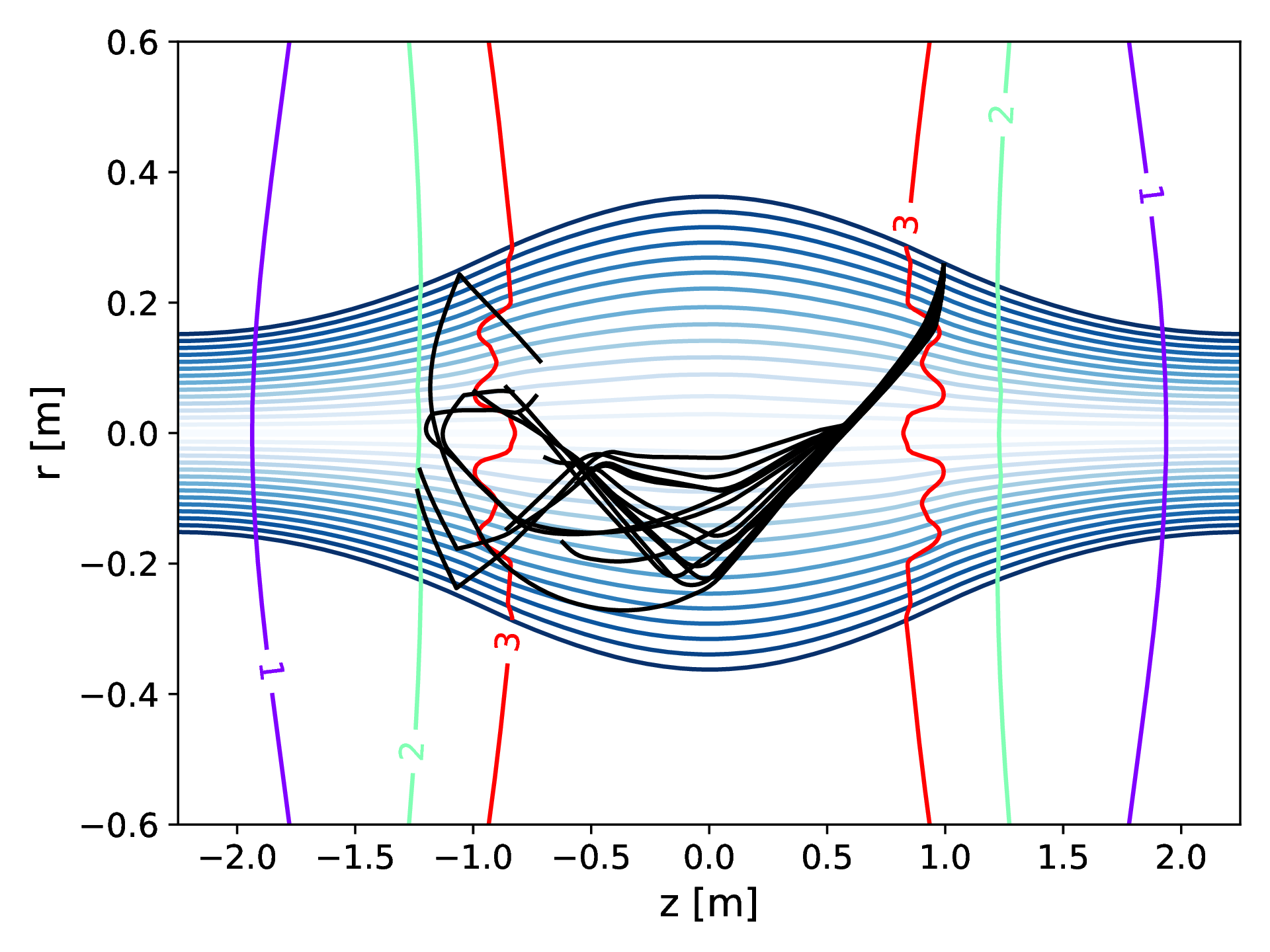}
  \caption{Plots of rays in black versus $r$ and $z$. Resonances are indicated with the rainbow colored labeled contours on the plot. Poloidal flux, $\psi$, surfaces are shown in blue.}
\label{fig:rf_rays}
\end{figure*}

The radial plasma $\langle E_i \rangle$ and $n$ profiles as well as the RF power deposition profiles from the simulations described above are shown in Figure~\ref{fig:rf_density}. The addition of RF heating did provide a slight increase in the overall particle confinement and particle density, but the overall plug performance remained well below that which could be obtained using $15~\mathrm{MW}$ of negative-ion NBI (reference the plot shown in Figure~\ref{fig:plug_dens} to see a comparable negative-ion NBI case). The majority of the improvement in particle confinement appears to have come from the $T_e$ increase due to the $\sim 1~\mathrm{MW}$ of electron directed RF heating by Landau damping. Ion directed heating's effect on plasma performance was a mixed. There was a significant amount of inverse cyclotron harmonic damping that diffused particles downwards in velocity-space into the loss cone. Furthermore, quasilinear diffusion cannot increase the density anywhere in velocity-space to a larger value than it is at the neutral beam source location. Thus, particles at the neutral beam source energy continued to dominate the overall velocity-space pitch-angle scattering rate. Energy confinement increased, with the plasma reaching high average ion energies due to the formation of an RF driven energetic ion tail and quasilinear diffusion driven pumpout of lower energy ions, however, this is actually undesirable in an end plug. Increasing average end plug particle energy without improving end plug particle confinement will not improve the tandem mirror central cell's performance and will only serve to drive up $\beta$ putting the end plug closer to pressure driven instability limits. One potential solution would be to use higher harmonics which bias the magnitude of the quasilinear diffusion operator upwards in velocity space due to the quasilinear diffusion operator's Bessel function dependence $Q \propto \mathrm{J}_n^2 ( k_\perp v_\perp/ n\Omega_i)$. However, harmonics greater than $n=3$ were found to be weakly damped in mirror end plug plasmas. Weak damping is undesirable in reactor scenarios as it can lead to poor antenna coupling and large in vessel fields that generate impurity sources. To increase damping rates, lower harmonics, $n=2,3$, had to be used. At lower harmonics, after the fast ion tail built up, damping became relatively strong, as shown in Figure~\ref{fig:rf_rays}, and the rays in the simulation damped after two passes through the resonance. 

As a result of simulations like the representative case shown here, RF ion heating was dropped from this study. The WHAM RF ion heating scheme was determined to not be scalable to tandem mirror fusion reactor end plugs, but will still be pursued experimentally on WHAM to quantify its effect on plasma performance and compared to simulations. Positive-ion NBI cannot effectively fuel a tandem mirror end plug at relevant values of $T_e$ and $\phi_p$ and RF quasilinear diffusion at $n = 2, 3$ harmonics with this scheme was not effective enough at increasing particle confinement times. Finally, in cases of both weak and the strong damping there could potentially be validity issues with the raytracing approximation here. In cases of weak damping, interference and diffraction may impact the wave power deposition, and in cases of strong damping, the distance between the antenna and the wave resonance are comparable to the wavelength of the wave. Because of this, full-wave simulations may be needed to adequately model RF in magnetic mirrors due to the breakdown of the raytracing approximation. Future work may revisit RF ion heating in conjunction with negative ion neutral beams, modified antenna launch spectra which optimize the shape of quasilinear diffusion operator and the use of a full wave code such as AORSA \citep{Jaeger2002}.

\bibliographystyle{jpp}

\bibliography{milestone5}

\end{document}